\documentclass[lettersize,journal]{IEEEtran}
\usepackage{amsmath,amsfonts}
\usepackage{algorithmic}
\usepackage{algorithm}
\usepackage{cuted}
\usepackage{array}
\usepackage[caption=false,font=normalsize,labelfont=sf,textfont=sf]{subfig}
\usepackage{textcomp}
\usepackage{stfloats}
\usepackage{url}
\usepackage{verbatim}
\usepackage{graphicx}
\usepackage{cite}
\usepackage{bm}
\begin{document}

\title{Free Space Optical Integrated Sensing and Communication Based on DCO-OFDM:\\ Performance Metrics and Resource Allocation}

\author{Yunfeng~Wen,
Fang~Yang,~\IEEEmembership{Senior~Member,~IEEE},
Jian~Song,~\IEEEmembership{Fellow,~IEEE},
and~Zhu~Han,~\IEEEmembership{Fellow,~IEEE}
\thanks{Part of this paper has been submitted to ICC 2024~\cite{OFDM_ISAC_ICC}. This work was supported in part by National Key Research and Development Program of China under Grant 2022YFE0101700; and in part by Beijing National Research Center for Information Science and Technology under Grant BNR2022RC01017. \emph{(Corresponding author: Fang~Yang.)}}
\thanks{Yunfeng~Wen and Fang~Yang are with the Department of Electronic Engineering, Tsinghua University, Beijing 100084, P. R. China, and also with the Key Laboratory of Digital TV System of Shenzhen City, Research Institute of Tsinghua University in Shenzhen, Shenzhen 518057, P. R. China (e-mail: wenyf22@mails.tsinghua.edu.cn; fangyang@tsinghua.edu.cn).}
\thanks{Jian Song is with the Department of Electronic Engineering, Tsinghua University, Beijing 100084, P. R. China, and also with the Shenzhen International Graduate School, Tsinghua University, Shenzhen 518055, P. R. China (e-mail: jsong@tsinghua.edu.cn).}
\thanks{Zhu Han is with the Department of Electrical and Computer Engineering, University of Houston, Houston, TX 77004 USA, and also with the Department of Computer Science and Engineering, Kyung Hee University, Seoul 446-701, South Korea (e-mail: hanzhu22@gmail.com).}
}

\maketitle

\begin{abstract}
  As one of the six usage scenarios of the sixth generation (6G) mobile communication system, integrated sensing and communication (ISAC) has garnered considerable attention, and numerous studies have been conducted on radio-frequency (RF)-ISAC. Benefitting from the communication and sensing capabilities of an optical system, free space optical (FSO)-ISAC becomes a potential complement to RF-ISAC. In this paper, a direct-current-biased optical orthogonal frequency division multiplexing (DCO-OFDM) scheme is proposed for FSO-ISAC. To derive the spectral efficiency for communication and the Fisher information for sensing as performance metrics, we model the clipping noise of DCO-OFDM as additive colored Gaussian noise to obtain the expression of the signal-to-noise ratio. Based on the derived performance metrics, joint power allocation problems are formulated for both communication-centric and sensing-centric scenarios. In addition, the non-convex joint optimization problems are decomposed into sub-problems for DC bias and subcarriers, which can be solved by block coordinate descent algorithms. Furthermore, numerical simulations demonstrate the proposed algorithms and reveal the trade-off between communication and sensing functionalities of the OFDM-based FSO-ISAC system.
\end{abstract}



\section{Introduction}\label{introduction}
Integrated sensing and communication (ISAC) is a technology that combines communication and sensing functionalities within an individual system~\cite{liu_ISAC_Overview}. Leveraging the coexistence, cooperation, and co-design of these functionalities, ISAC can efficiently utilize the limited resources of power, spectrum, and hardware to provide both high-rate communication capacities and high-precision sensing abilities simultaneously~\cite{chiriyath_RCC}. Therefore, ISAC is recognized as one of the six usage scenarios for the sixth generation (6G) mobile communication system and also a key enabler for future applications like the Internet of Things (IoT)~\cite{cui_ISAC_IoT}, intelligent transportation systems~\cite{ma_JRC_vehicle}, environmental monitoring~\cite{chatzitheodoridi_SAR_Communication}, and human-computer interaction~\cite{lien_Soli_Gesture}. With its revolutionary potential, ISAC has garnered considerable attention from both academia and industry, and numerous studies have been conducted spanning several areas like waveform design~\cite{zhang_LFM-CPM,wei_ISAC_Signals_Survey}, networking~\cite{peng_SC_Network}, resource allocation~\cite{wang_Achieving_Performance_Bounds,zhao_ResourceAllocation_ISAC}, and information theory~\cite{chiriyath_Inner_Bounds}.

Despite the rapid development witnessed by radio-frequency (RF)-ISAC, free space optical (FSO)-ISAC has received much less attention. However, the optical spectrum offers \emph{three advantages} in the context of ISAC. First, FSO communication can utilize the large unlicensed optical spectrum to achieve high data rate~\cite{khalighi_Survey_FSO_Information_Theory}. Second, due to the large bandwidth and narrow beams of lasers, FSO sensing devices like light detection and ranging (LiDAR) can achieve both high distance resolution and high angle resolution~\cite{li_Lidar_Autonomous_Driving}. Third, FSO communication and FSO sensing are free from electromagnetic interference thanks to their line-of-sight (LoS) channels and non-penetration properties~\cite{gailani_Survey_FSO}. Moreover, well-established ISAC waveforms can be utilized for FSO-ISAC by subcarrier intensity modulation (SIM)~\cite{liu_PPM_MSK_SIM}, and the intensity modulation and direct detection (IM/DD) scheme can be adopted by both FSO communication and FSO sensing~\cite{mohamed_IMDD_Performance}. Therefore, with joint design and optimization, FSO-ISAC can also become a powerful complement to RF-ISAC.


Waveform design is a fundamental problem for FSO-ISAC, and various waveforms have been proposed to enable simultaneous communication and sensing. For instance, the boomerang transmission system combines automotive LiDAR with FSO communication, which employs time-hopping pulses as the ISAC signal~\cite{suzuki_LaserRadar_Visible}. Similarly, a group of optical pulses is utilized for FSO-ISAC by the pulse sequence sensing and pulse position modulation scheme, which exploits pulse positions for both communication and sensing~\cite{wen_PSSPPM_ISAC}. For continuous-wave FSO-ISAC, a unified waveform is proposed based on quadrature phase shift keying and direct sequence spread spectrum, which loads communication data on a pseudorandom noise sequence~\cite{cao_Unified_Optcial_ISAC}. In addition, a frequency-modulated continuous-wave coherent LiDAR is addressed to provide downlink communications capability~\cite{xu_FMCW_Lidar}. Moreover, the phase-shift laser ranging with communication is investigated to mitigate ambiguity caused by the phase-coded signal~\cite{hai_Remote_PS_LiDAR}. However, the majority of existing FSO-ISAC waveforms focus on pulses or single-carrier continuous waves, with limited attention paid to the multi-carrier scheme, which restricts the achievable data rate of FSO-ISAC.

Among widely utilized multi-carrier schemes, orthogonal frequency division multiplexing (OFDM) has gained popularity in RF-ISAC due to its time-frequency interpretation~\cite{baquero_OFDM_LET_5GNR}. On one hand, signal processing techniques have been thoroughly studied for both communication and sensing based on OFDM. For instance, the maximum-likelihood method can be utilized to approach the Cramèr-Rao Bound (CRB) asymptotically for point targets in a prior, at the expense of a deteriorated ambiguity function~\cite{sturm_OFDM_JRC}. Towards this end, an element-wise-division method can be adopted for OFDM-based ISAC to achieve a higher peak-to-sidelobe ratio~\cite{sturm_Waveform_Design}. On the other hand, plentiful joint optimization techniques arise from novel performance metrics. Apart from the conventional CRB metric~\cite{liyanaarachchi_Opt_Waveform}, mutual information and power minimization present novel objectives for OFDM-based ISAC, with constraints on the quality of communication and sensing services~\cite{liu_Adaptive_OFDM_IRC,zhu_Power_Minimization}. Furthermore, considering practical limitations in capturing a precise channel response, an uncertainty model can be exploited to obtain robust solutions~\cite{shi_Robust_OFDM}. Therefore, OFDM is a promising candidate for FSO-ISAC thanks to its high spectral efficiency and mature signal processing techniques.

Nevertheless, a conventional OFDM waveform for RF-ISAC is not readily applicable to FSO-ISAC, since an optical waveform based on IM/DD is restricted to being real and non-negative. Leveraging the Hermitian symmetry in the frequency domain, researchers have proposed optical OFDM schemes like direct-current-biased optical OFDM (DCO-OFDM), asymmetrically clipped optical OFDM (ACO-OFDM), and asymmetrically clipped direct-current-biased optical OFDM (ADO-OFDM)~\cite{dissanayake_Comparison_OFDM, huang_ADOOFDM_Allocation}. As non-negative clipping is generally involved in optical OFDM, the effects of non-linear distortion should be considered either in the performance evaluation or in the detection~\cite{tan_SequenceDetection_DCOOFDM, marmin_Recovery_Clipped}. Although the Bussgang theorem has been introduced to model the clipping noise~\cite{dimitrov_Clipping_OFDM}, it may fail in an ISAC scenario where the power is not uniformly distributed on each subcarrier. Furthermore, the complicated expression of the clipping noise also hampers the optimization of FSO-ISAC~\cite{jiang_ClippingNoise}.

\begin{figure*}[tp]
  \centering
  \includegraphics[width=0.96\textwidth]{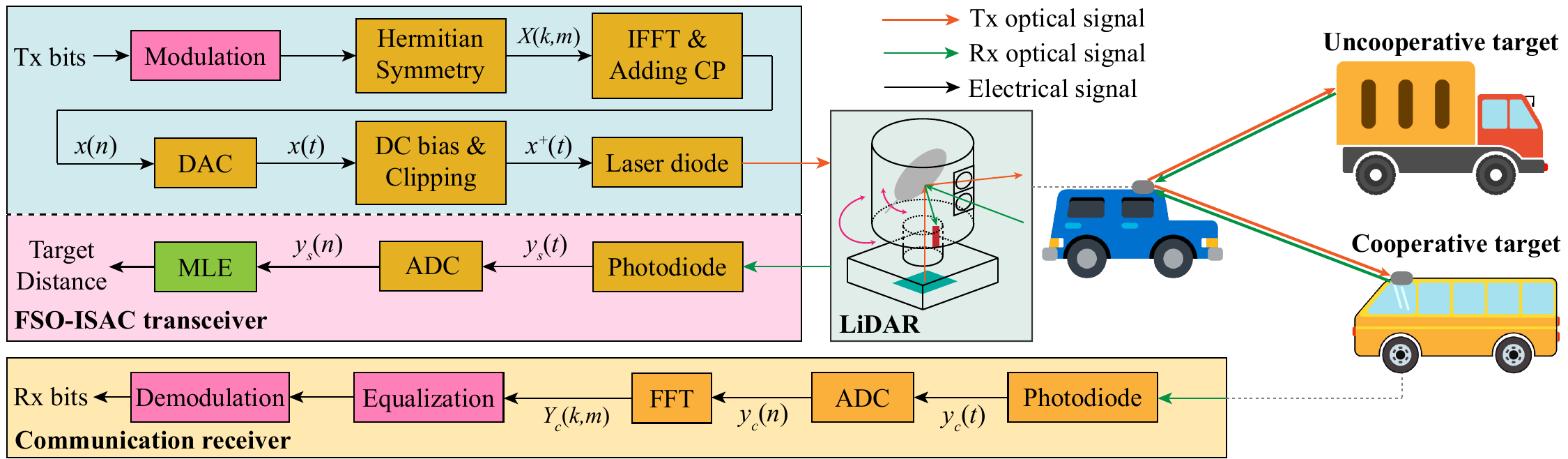}
  \caption{System model of the proposed DCO-OFDM scheme for FSO-ISAC. ADC: analog-to-digital converter; DAC: digital-to-analog converter; FFT: fast Fourier transform; IFFT: inverse fast Fourier transform; CP: cyclic prefix.}
  \label{fig:system_model}
\end{figure*}

Motivated by the potential and challenges mentioned above, we propose a DCO-OFDM scheme for FSO-ISAC in this paper, which aims at achieving both a higher spectral efficiency and a superior sensing precision. Specifically, our contributions are summarized as follows:
\begin{itemize}
  \item First, the system model of the DCO-OFDM scheme for FSO-ISAC is presented to provide simultaneous communication and sensing capabilities for a LiDAR. Based on the established system model, we describe the working principles and derive the performance metrics for both communication and sensing subsystems. In addition, considering the uneven power allocation on subcarriers, the clipping noise is modelled as additive colored Gaussian noise instead of the conventional white noise model, which yields a more precise expression for the signal-to-noise ratio (SNR).
  \item Second, given the performance metrics of both communication and sensing, the power allocation problems are formulated for both communication-centric and sensing-centric scenarios. To solve these non-convex problems, the joint power allocation problem is decomposed into two sub-problems, i.e., a sub-problem for direct-current (DC) bias and a sub-problem for power allocation on subcarriers. In consequence, the sub-problem for DC bias degenerates into a single-variable problem, while the sub-problem for power allocation on subcarriers is convex. 
  \item Third, a block coordinate descent (BCD) algorithm is proposed to solve the joint power allocation problem, which alternately solves the sub-problems for DC bias and power allocation on subcarriers. Specifically, the sub-problem for DC bias can be solved by the golden search algorithm, while a closed-form expression of power allocation on subcarriers can be derived. Furthermore, an iterative optimization algorithm is proposed to attain the dual variables in the expression of power allocation on subcarriers. Numerical simulations demonstrate the proposed algorithms and reveal the trade-off between communication and sensing functionalities.
\end{itemize}

The rest of this paper is organized as follows. In Section~\ref{system_model}, the system model is introduced, and the performance metrics for both communication and sensing are derived. Section~\ref{clipping_noise_statistics} presents the model of non-negative clipping, where the clipping noise is modelled as additive colored Gaussian noise. In Section~\ref{power_allocation_c}, the communication-centric power allocation problem is formulated and decomposed, with a BCD algorithm proposed to solve the joint power allocation problem. Similarly, the sensing-centric power allocation problem is investigated in Section~\ref{power_allocation_s}. Moreover, detailed simulation results are illustrated in Section~\ref{numerical_simulation}, and the conclusion is drawn in Section~\ref{conclusion}.


\section{System Model}\label{system_model}
In this section, we introduce the system model of the DCO-OFDM scheme for FSO-ISAC. As illustrated in Fig.~\ref{fig:system_model}, the FSO-ISAC transceiver is based on a LiDAR, which provides active sensing abilities for both cooperative and uncooperative targets. Moreover, once the communication receiver of the cooperative target is not obstructed, a communication link can be established to achieve sensing-communication integration. The DCO-OFDM signal model and channel model are introduced in Subsections~\ref{signal_model} and~\ref{channel_model}, respectively. Subsequently, the communication with the cooperative target is investigated in Subsection~\ref{comm_subsystem}, while the active sensing process is elaborated in Subsection~\ref{sensing_subsystem}.

\subsection{DCO-OFDM Signal Model}\label{signal_model}
For optical OFDM, the time-domain signal is constrained to be real and non-negative. Therefore, the Hermitian symmetry is imposed to the frequency-domain subcarriers, i.e., $X\left(k,m\right)=X^*\left(N-k,m\right)$, where $X\left(k,m\right)$ is transmitted on the $k$-th subcarrier of the $m$-th OFDM symbol. In addition, $X\left(0,m\right)$ and $X\left(N/2,m\right)$ are set to zeros, yielding $N/2-1$ independent subcarriers in each OFDM symbol. Then, the time-domain OFDM signal is expressed as
\begin{equation}
  \begin{split}
    x\left(t\right)=&\frac{1}{\sqrt{N}}\sum_{m=0}^{M-1}{\left\{\sum_{k=0}^{N-1}\left[X\left(k,m\right)\right.\right.}\\
    &{\left.\left.\exp\left(j2\pi k\Delta f\left(t-mT_o\right)\right)\right]\text{rect}\left(\frac{t-mT_o}{T_o}\right)\right\}},
  \end{split}
\end{equation}

{\noindent}where notations $M$, $N$, and $\Delta f$ are the amount of OFDM symbols in each frame, the number of subcarriers, and the subcarrier spacing, respectively. Besides, a guard interval is concatenated in front of each OFDM symbol to avoid inter-symbol interference. The durations of the elementary OFDM symbol, the guard interval, and the total OFDM symbol are denoted as $T$, $T_g$, and $T_o$, respectively, where $T=1/\Delta f$ and $T_o=T+T_g$. In addition, the occupied bandwidth of $x\left(t\right)$ is $B=N\Delta f$, and $\text{rect}\left(\cdot\right)$ denotes a rectangular window, during which the signal is sampled at the rate of $R_s=N/T$ for subsequent digital signal processing.

To obtain a non-negative signal, a DC component $b$ is added to $x\left(t\right)$ to generate the biased signal. Then, the negative part of the biased signal is clipped, which yields the transmitted DCO-OFDM signal as
\begin{equation}\label{Clipping}
  x^{+}\left(t\right)=\left\{x\left(t\right)+b\right\}^+,
\end{equation}

{\noindent}where the notation $\left\{\cdot\right\}^+$ is defined as $\left\{x\right\}^+=\max\left\{x,0\right\}$.

\subsection{Channel Model}\label{channel_model}
For terrestrial scenarios, the FSO signal propagates through an atmospheric channel, where propagation impairments originate from atmospheric attenuation, turbulence, geometric loss, etc. Therefore, the channel response $h\left(t\right)$ can be decomposed into a stationary channel gain $\bar{h}$ and a normalized impulse response $\tilde{h}\left(t\right)$ corresponding to the dispersion. In consequence, the channel response is written as
\begin{equation}\label{channel_response}
  h\left(t\right)=\bar{h}\cdot\tilde{h}\left(t\right)=L_aL_tL_gG_TG_R\tilde{h}\left(t\right),
\end{equation}

{\noindent}where $L_a$, $L_t$, $L_g$, $G_T$, and $G_R$ denote the atmospheric attenuation, the scintillation brought by turbulence, the geometric loss, the gain of the transmitter, and the gain of the receiver, respectively. $G_T$ and $G_R$ can be derived by the gain of optical amplifiers and the receiver responsivity, while the extinctions $L_t$, $L_a$, and $L_g$ are discussed as follows.

The log-normal distribution is adopted to model the FSO turbulent channel under weak atmospheric turbulence, and the scintillation term in~(\ref{channel_response}) follows the distribution of
\begin{equation}
  p\left(L_t\right)=\frac{1}{L_t\sqrt{2\pi\sigma_t^2}}\exp\left(-\frac{1}{2\sigma_t^2}{\left(\ln\dfrac{L_t}{\bar{L}_t}+\dfrac{\sigma_t^2}{2}\right)}^2\right),
\end{equation}

{\noindent}where $\bar{L}_t$ and $\sigma_t^2$ denote the average turbulence power and the scintillation index, respectively. In this paper, $\bar{L}_t$ is normalized, and $\sigma_t^2$ can be obtained by the Rytov approximation as
\begin{equation}
  \sigma_t^2\approx 1.23{\left(\frac{2\pi}{\lambda}\right)}^{7/6}L^{11/6}C_n^2,
\end{equation}

{\noindent}where $C_n^2$, $\lambda$, and $L$ denote the refractive index, the optical wavelength, and the optical path length, respectively.

The atmospheric attenuation arises from the absorption and scattering of fog, haze, and heavy rain, and an empirical formula to calculate the attenuation is given by
\begin{equation}
  L_a=\alpha V^{\beta}L,
\end{equation}

{\noindent}where $V$ denotes the atmospheric visibility~\cite{nebuloni_FSO_Visibility}. Besides, the coefficients $\alpha$ and $\beta$ can be determined by the optical wavelength and the visibility range.

The geometric loss arises from the divergence of laser beams between the transmitter and the receiver. Supposing that perfect alignment is achieved by the acquisition, tracking, and pointing (ATP) mechanism, the geometric loss can be expressed as
\begin{equation}
  L_g=\frac{A}{\pi{\left(\frac{L\theta}{2}\right)}^2},
\end{equation}

{\noindent}where $A$ and $\theta$ denote the receiver aperture and the beam divergence angle, respectively.


\subsection{Communication Subsystem}\label{comm_subsystem}
The received communication signal is expressed as
\begin{equation}
  y_c\left(n\right)=x_c\left(\frac{n}{R_s}\right)+w_c\left(n\right),
\end{equation}

{\noindent}where thermal noise and shot noise are modelled as additive white Gaussian noise (AWGN), i.e., $w_c\left(n\right)\sim\mathcal{N}\left(0,\sigma_c^2\right)$, with a power spectral density (PSD) of $N_c=\sigma_c^2/B$. Besides, the received signal without noise is given by
\begin{equation}
  x_c\left(t\right)=\bar{h}_c\cdot x^+\left(t\right)*\tilde{h}_c\left(t\right),
\end{equation}

{\noindent}where the operator $*$ denotes the linear convolution, while $\bar{h}_c$ and $\tilde{h}_c\left(t\right)$ denote the stationary channel gain and the channel response for communication, respectively. If the delay spread of $h_c\left(t\right)$ does not exceed the duration of the guard interval, the linear convolution can be substituted by the cyclic convolution, and a frequency-domain expression for the communication signal can be obtained as
\begin{equation}
  Y_c\left(k,m\right)=X^{+}\left(k,m\right)\tilde{H}_c\left(k\right)+\tilde{W}_c\left(k\right),
\end{equation}

{\noindent}where $X^{+}\left(k,m\right)$ and $\tilde{H}_c\left(k\right)$ are the discrete Fourier transform (DFT) of $x^{+}\left(n/R_s\right)$ and $\tilde{h}_c\left(n/R_s\right)$, respectively. Besides, the impact of $\bar{h}_c$ is normalized into the noise term, i.e.,
\begin{equation}
  \tilde{W}_c\left(k\right)=\frac{1}{\bar{h}_c}\sum_{n=0}^{N-1}{w_c\left(n\right)\exp\left(-\frac{j2\pi nk}{N}\right)}.
\end{equation}

Based on the ergodic channel capacity derived in~\cite{sharma_Capacity_FSO_Turbulence}, the performance metric of the communication subsystem, i.e., spectral efficiency, can be expressed as
\begin{equation}\label{Capacity}
  C\left(b,\tilde{P}\left(k\right)\right)=\frac{1}{BT_o}\sum_{k=1}^{\frac{N}{2}-1}{\log\left(1+\gamma_c\left(k\right)\tilde{P}\left(k\right)\right)},
\end{equation}

{\noindent}where $\gamma_c\left(k\right)$ denotes the normalized SNR on the $k$-th subcarrier for the communication signal, and $\tilde{P}\left(k\right)$ is the normalized power allocation for the $k$-th subcarrier.

Additionally, to derive the expression of $\tilde{P}\left(k\right)$, the total electrical power is written as $P$, which is utilized by both the DC bias and the communication signal on each subcarrier, i.e.,
\begin{equation}
  P=b^2+\sum_{k=1}^{\frac{N}{2}-1}{\mathbb{E}\left({\lvert X\left(k,m\right) \rvert}^2\right)},
\end{equation}

{\noindent}where $\mathbb{E}\left(X\right)$ denotes the expectation of $X$.

Then, the normalized power allocation for the $k$-th subcarrier is given by
\begin{equation}
  \tilde{P}\left(k\right)=\dfrac{\mathbb{E}\left({\lvert X\left(k,m\right) \rvert}^2\right)}{P-b^2}.
\end{equation}


\subsection{Sensing Subsystem}\label{sensing_subsystem}
Similar to the communication subsystem, the received sensing signal is expressed as
\begin{equation}
  y_s\left(n\right)=\mathfrak{R}x_s\left(\frac{n}{R_s}-\tau_0\right)+w_s\left(n\right),
\end{equation}

{\noindent}where $w_s\left(n\right)\sim\mathcal{N}\left(0,\sigma_s^2\right)$ is AWGN with a PSD of $N_s=\sigma_s^2/B$. Additionally, $\tau_0$ and $\mathfrak{R}$ denote the time of flight (ToF) and the target reflectivity, respectively. Moreover, the received signal without noise is given by
\begin{equation}
  x_s\left(t\right)=\bar{h}_s\cdot x^+\left(t\right)*\tilde{h}_s\left(t\right),
\end{equation}

{\noindent}where $\bar{h}_s$ and $\tilde{h}_s\left(t\right)$ denote the stationary channel gain and the channel response for sensing, respectively.

Once the FSO signal is received, the sensing receiver can utilize the cross-correlation method to obtain the maximum-likelihood estimation (MLE) of $\tau_0$ as
\begin{equation}
  \hat{\tau}_0=\mathop{\mathrm{arg\ max}}\limits_{\tau}{\sum_{n=0}^{MT_oR_s-1}{y_s\left(n\right)x\left(\frac{n}{R_s}-\tau\right)}},
\end{equation}

{\noindent}which further yields the target distance as $\hat{L}=2\hat{\tau}_0/c$ with $c$ denoting the speed of light. To evaluate its performance metric, the following theorem can be utilized.

\textbf{Theorem 1: }\textit{The Fisher information of the target distance estimation based on an OFDM signal is expressed as}
\begin{equation}\label{Fisher_info_metric}
  I\left(b,\tilde{P}\left(k\right)\right)=\frac{8\pi^2M{\Delta f}^2}{N}\sum_{k=1}^{\frac{N}{2}-1}{k^2\gamma_s\left(k\right)\tilde{P}\left(k\right)},
\end{equation}

{\noindent}\textit{where $\gamma_s\left(k\right)$ is the normalized SNR on the $k$-th subcarrier for the sensing signal.}

\textit{Proof: }See Appendix~\ref{append_crb}.

With appropriate interpolation schemes, the precision of MLE approaches CRB asymptotically~\cite{kay_Fundamentals}, which is inversely proportional to the Fisher information. Therefore, the Fisher information in~(\ref{Fisher_info_metric}) is adopted as the performance metric for sensing.


\section{Clipping Noise Statistics}\label{clipping_noise_statistics}
As described in~(\ref{Clipping}), the transmitted signal $x^{+}\left(n\right)$ is a clipped version of $x\left(n\right)$, thus having distinct statistics from the original OFDM signal. Supposing that $X\left(k,m\right)$ obeys the Gaussian distribution independently, $x\left(n\right)$ is a Gaussian random process with a variance of $\sigma_x^2=\left(P-b^2\right)/N$ and is cyclostationary when $\mathbb{E}\left(X\left(k,m\right)\right)=0$.

Moreover, the Bussgang theorem indicates that $x^{+}\left(n\right)$ can be decomposed as
\begin{equation}
  x^{+}\left(n\right)=\mathcal{K}x\left(n\right)+w_{p}\left(n\right),
\end{equation}

{\noindent}where $w_{p}\left(n\right)$ is the clipping noise uncorrelated to $x\left(n\right)$~\cite{Bussgang_Theorem}. Denoting the complementary cumulative distribution function of the standard Gaussian distribution as $Q\left(\cdot\right)$, the coefficient $\mathcal{K}$ can be calculated as
\begin{equation}
  \mathcal{K}=\frac{\mathbb{E}\left(x\left(n\right)x^{+}\left(n\right)\right)}{\mathbb{E}\left(x^2\left(n\right)\right)}=Q\left(\lambda_b\right),
\end{equation}

{\noindent}where $\lambda_b=-b/\sigma_x$ is the normalized clipping level.

Furthermore, to obtain the statistics of the clipping noise, $w_{p}\left(n\right)$ is viewed as the output of a non-linear memoryless system $g_p\left(x\right)$ whose input is $x\left(n\right)$, i.e.,
\begin{equation}
  w_{p}\left(n\right)=\left\{x\left(n\right)+b\right\}^+-b-\mathcal{K}x\left(n\right):=g_{p}\left(x\left(n\right)\right).
\end{equation}

Assuming that $\mathbb{E}\left(X\left(k,m\right)\right)=0$, $w_{p}\left(n\right)$ is also a cyclostationary random process, and its auto-correlation function is defined as
\begin{equation}
  R_{w_{p}}\left(n-m\right)=\mathbb{E}\left(g_{p}\left(x\left(n\right)\right)g_{p}\left(x\left(m\right)\right)\right),
\end{equation}

{\noindent}which can be derived by the following theorem.

\textbf{Theorem 2: }\textit{the auto-correlation function of the clipping noise can be expressed as}
\begin{equation}\label{R_w_clip}
  R_{w_{p}}\left(n\right)=\mathcal{I}\left(r\right)+C_1r+C_2,
\end{equation}

{\noindent}\textit{where the integral is defined as}
\begin{equation}\label{Price_integral}
  \mathcal{I}\left(r\right)=\int_{-\sigma_x^2}^{r}{\int_{-\sigma_x^2}^{s}{\left[\frac{1}{2\pi\sqrt{\sigma_x^4-t^2}}\exp\left(-\frac{b^2}{\sigma_x^2+t}\right)\right]dt}ds},
\end{equation}

{\noindent}\textit{and $r=R_x\left(n\right)$ is the auto-correlation function of the original OFDM signal $x\left(n\right)$. Besides, the constants $C_1$ and $C_2$ are calculated as}
\begin{subequations}
    \begin{align}
      C_1=&\frac{1}{\sigma_x^2}\left(\mathbb{E}\left(\left(w_{p}\left(n\right)\right)^2\right)-C_2-\mathcal{I}\left(\sigma_x^2\right)\right),\label{Ya}\\
      C_2=&\left(\mathbb{E}\left(w_{p}\left(n\right)\right)\right)^2-\mathcal{I}\left(0\right).\label{Yb}
    \end{align}
\end{subequations}

\textit{The first-order and second-order momentums of $w_{p}\left(n\right)$ in (\ref{Ya}) and (\ref{Yb}) are given by}
\begin{subequations}
  \begin{align}
    \mathbb{E}\left(w_{p}\left(n\right)\right)=&\lvert \sigma_x\left(\lambda_b\left(1-Q\left(\lambda_b\right)\right)-\phi\left(\lambda_b\right)\right) \rvert,\\
    \mathbb{E}\left(w_{p}\left(n\right)\right)^2=&\sigma_x^2\left(\lambda_b^2\left(1-Q\left(\lambda_b\right)\right)\right.\notag\\
    &\left.+\lambda_b\phi\left(\lambda_b\right)+Q\left(\lambda_b\right)-\mathcal{K}^2\right),
  \end{align}
\end{subequations}

{\noindent}\textit{where $\phi\left(x\right)$ is the probability distribution function of the standard Gaussian distribution.}

\textit{Proof: }See Appendix~\ref{append_clipping_corr}.

Subsequently, the PSD of clipping noise is obtained by the Wiener-Khinchin theorem as
\begin{equation}
  P_{w_{p}}\left(k\right)=\sum_{n=0}^{N-1}{R_{w_{p}}\left(n\right)\exp\left(-\frac{j2\pi nk}{N}\right)}.
\end{equation}

Consequently, by modelling the frequency-domain clipping noise as a Gaussian random process uncorrelated to the original signal, the normalized SNRs of the $k$-th subcarrier for communication and sensing are expressed respectively as
\begin{subequations}
  \begin{align}
    &\gamma_c\left(k\right)=\dfrac{{\lvert\tilde{H}_c\left(k\right)\rvert}^2\mathcal{K}^2\left(P-b^2\right)}{\dfrac{N_c\Delta f}{2\mathbb{E}\left(\bar{h}_c\right)^2}+{\lvert\tilde{H}_c\left(k\right)\rvert}^2P_{w_{p}}\left(k\right)},\label{snr_c}\\
    &\gamma_s\left(k\right)=\dfrac{{\lvert\tilde{H}_s\left(k\right)\rvert}^2\mathcal{K}^2\left(P-b^2\right)}{\dfrac{N_s\Delta f}{2\mathfrak{R}^2\mathbb{E}\left(\bar{h}_s\right)^2}+{\lvert\tilde{H}_s\left(k\right)\rvert}^2P_{w_{p}}\left(k\right)},\label{snr_s}
  \end{align}
\end{subequations}

{\noindent}where $\tilde{H}_s\left(k\right)$ is the DFT of $\tilde{h}_s\left(n/R_s\right)$.


\section{Communication-centric Resource Allocation}\label{power_allocation_c}
As indicated by~(\ref{snr_c}) and (\ref{snr_s}), the normalized SNR depends on the power allocation for DC bias and subcarriers, which affects both the spectral efficiency and the Fisher information. Therefore, the DC bias $b$ and power allocation $\tilde{P}\left(k\right)$ for subcarriers should be optimized jointly to achieve superior performance metrics. In this section, we consider the joint power allocation problem for a communication-centric scenario, which is first formulated in Subsection~\ref{power_allocation_c_joint}. Then, the joint power allocation problem is decomposed into a sub-problem for DC bias and a sub-problem for power allocation on subcarriers, which are elaborated in Subsections~\ref{power_allocation_c_bias} and~\ref{power_allocation_c_subcarrier}, respectively. Furthermore, the convergence of proposed algorithms is investigated in Subsection~\ref{converge_c}, and their computational complexity is given in Subsection~\ref{complexity_c}.


\subsection{Joint Power Allocation Problem}\label{power_allocation_c_joint}
The joint power allocation problem achieves the optimal communication performance and guarantees the sensing performance simultaneously, which is formulated as
\begin{subequations}\label{opt_c}
  \begin{align}
    &\text{(P1):}& &\mathop{\max}\limits_{b,\tilde{P}\left(k\right)}\ & &C\left(b,\tilde{P}\left(k\right)\right),&\label{opt_c_obj}\\
    & & &\text{s.t.}\ & &I\left(b,\tilde{P}\left(k\right)\right)\geq\varsigma_0^2,&\label{opt_c_constr_sensing}\\
    & & &\ & &0\leq b\leq\sqrt{P},&\label{opt_c_constr_bias}\\
    & & &\ & &0\leq\tilde{P}\left(k\right)\leq \tilde{P}_m,\label{opt_c_constr_power_each}&\\
    & & &\ & &\sum_{k=1}^{\frac{N}{2}-1}{\tilde{P}\left(k\right)}=\frac{1}{2},&\label{opt_c_constr_power_total}
  \end{align}
\end{subequations}

\begin{figure}[!t]
  \begin{algorithm}[H]
      \caption{BCD Algorithm for Communication-centric Scenario}
      \label{alg:opt_alter_c}
      \begin{algorithmic}[1]
          \renewcommand{\algorithmicrequire}{\textbf{Input:}}
          \renewcommand{\algorithmicensure}{\textbf{Output:}}
          \REQUIRE Tolerance $\epsilon_c$. Initial solution $b^{\left(0\right)}$ and $\tilde{P}^{\left(0\right)}\left(k\right)$.
          \ENSURE Optimal $b_c^*$ and $\tilde{P}_c^*\left(k\right)$ to~(P1).
          \STATE $i\leftarrow 0$, $C^{\left(0\right)}\leftarrow 0$.
          \WHILE{$\lvert C^{\left(i+1\right)}-C^{\left(i\right)} \rvert\geq\epsilon_c$}
            \STATE Given $\tilde{P}^{\left(i\right)}\left(k\right)$, solve (P1-1) to obtain $b^{\left(i+1\right)}$.
            \STATE Given $b^{\left(i+1\right)}$, solve (P1-2) to obtain $\tilde{P}^{\left(i+1\right)}\left(k\right)$ and $C^{\left(i+1\right)}$.
            \STATE $i\leftarrow i+1$.
          \ENDWHILE
          \STATE $b_c^*\leftarrow b^{\left(i\right)}$, $\tilde{P}_c^*\left(k\right)\leftarrow \tilde{P}^{\left(i\right)}\left(k\right)$.
      \end{algorithmic}
  \end{algorithm}
\end{figure}

{\noindent}where $c/2\varsigma_0$ is the desired precision of sensing, and $\tilde{P}_m$ is the maximum normalized power that can be allocated to an individual subcarrier. The constraint (\ref{opt_c_constr_sensing}) guarantees that the sensing subsystem can achieve the desired precision, while constraints (\ref{opt_c_constr_bias}) and (\ref{opt_c_constr_power_each}) restrict $b$ and $\tilde{P}\left(k\right)$ to be non-negative, respectively. Additionally, constraints (\ref{opt_c_constr_bias}) and (\ref{opt_c_constr_power_total}) impose restrictions on the total power, and we set $\tilde{P}_m<1/2<\left(N/2-1\right)\tilde{P}_m$ to avoid invalid constraints.

Unfortunately, the joint optimization for $b$ and $\tilde{P}\left(k\right)$ is non-convex in~(P1). By decomposing~(P1) into an optimization problem for $b$ with fixed $\tilde{P}\left(k\right)$ and an optimization problem for $\tilde{P}\left(k\right)$ with fixed $b$, we obtain a single-variable non-convex optimization problem and a convex optimization problem with $N/2-1$ variables, respectively. However, as indicated by (\ref{R_w_clip}), $\tilde{P}\left(k\right)$ affects the PSD of the clipping noise, and thus alters the optimal $b$. On the other hand, $b$ in turn affects the PSD of clipping noise, and the optimal $\tilde{P}\left(k\right)$ is also changed. Therefore, the joint power allocation problem (P1) cannot be solved properly in a single round.

Towards this end, we adopt a BCD algorithm to solve (P1). Given the power allocation $\tilde{P}^{\left(i\right)}\left(k\right)$ in the feasible region of~(\ref{opt_c_constr_power_each}) and (\ref{opt_c_constr_power_total}), the DC bias is optimized to obtain $b^{\left(i+1\right)}$. Then, the power allocation is further optimized to obtain $\tilde{P}^{\left(i+1\right)}\left(k\right)$ with the given $b^{\left(i+1\right)}$. The iteration repeats until the objective $C^{\left(i\right)}$ converges, which is summarized in~\textbf{Algorithm~\ref{alg:opt_alter_c}}. Since its convergence cannot be derived theoretically, the iteration of~\textbf{Algorithm~\ref{alg:opt_alter_c}} should be aborted if the value of $C^{\left(i\right)}$ diverges.


\subsection{Power Allocation for DC Bias}\label{power_allocation_c_bias}
The sub-problem for DC bias is formulated as
\begin{subequations}\label{opt_b_c}
  \begin{align}
    &\text{(P1-1):}& &\mathop{\max}\limits_{b}\ & &C\left(b,\tilde{P}\left(k\right)\right),&\\
    & & &\text{s.t.}\ & &0\leq b\leq\sqrt{P}.&
  \end{align}
\end{subequations}

To maximize the spectral efficiency $C$, the log-sum of the normalized SNR $\gamma_c\left(k\right)$ should be maximized. Although the DC bias $b$ affects $\mathcal{K}$, $P_{w_p}\left(k\right)$, and $P-b^2$ simultaneously, $\gamma_c\left(k\right)$ is generally dominated by the numerator in~(\ref{snr_c}) for large $N_c$, i.e.,
\begin{equation}\label{gamma_c}
  \begin{split}
    &\gamma_c\left(k\right)\approx\frac{2{\left(\mathbb{E}\left(\bar{h}_c\right)\right)}^2{\lvert\tilde{H}_c\left(k\right)\rvert}^2\mathcal{K}^2\left(P-b^2\right)}{N_c\Delta f},\\
    &\text{if}\ \frac{N_c\Delta f}{2{\left(\mathbb{E}\left(\bar{h}_c\right)\right)}^2}\gg{\lvert\tilde{H}_c\left(k\right)\rvert}^2P_{w_p}\left(k\right).
  \end{split}
\end{equation}

{\noindent}Consequently, $\gamma_c\left(k\right)$ is a unimodal function of $b$, and the golden search method can be adopted to obtain the optimal solution $b_c^*$ for the communication-centric scenario.


\subsection{Power Allocation for Subcarriers}\label{power_allocation_c_subcarrier}
Once $b_c^*$ is obtained by solving~(P1-1), the normalized power allocation $\tilde{P}\left(k\right)$ for subcarriers is further optimized to maximize the spectral efficiency. Meanwhile, for the sake of conciseness, the clipping noise PSD $P_{w_p}\left(k\right)$ is assumed to be a constant during the optimization. Therefore, the sub-problem is formulated as
\begin{subequations}\label{opt_Pk_c}
  \begin{align}
    &\text{(P1-2):}& &\mathop{\max}\limits_{\tilde{P}\left(k\right)}& &C\left(b,\tilde{P}\left(k\right)\right),&\\
    & & &\text{s.t.}\ & &I\left(b,\tilde{P}\left(k\right)\right)\geq\varsigma_0^2,&\label{opt_c_constr_sensing_2}\\
    & & &\ & &\sum_{k=1}^{\frac{N}{2}-1}{\tilde{P}\left(k\right)}=\frac{1}{2},&\\
    & & &\ & &0\leq\tilde{P}\left(k\right)\leq \tilde{P}_m.&
  \end{align}
\end{subequations}

As indicated by~(\ref{Capacity}) and~(\ref{Fisher_info_metric}), the objective of~(P1-2) is a concave function of $\tilde{P}\left(k\right)$, while the constraints are affine functions of $\tilde{P}\left(k\right)$. In consequence, maximizing a concave objective under affine constraints yield a convex optimization problem. Therefore, the optimal solution to~(P1-2) is obtained if and only if Karush-Kuhn-Tucker (KKT) conditions hold, based on which three cases can be considered as follows.

\subsubsection{Case A}The sensing constraint (\ref{opt_c_constr_sensing_2}) is first assumed to be inactive, and then the power allocation problem can be solved by the conventional water-filling method. Therefore, the normalized power allocation for the $k$-th subcarrier is expressed as
\begin{subequations}\label{opt_Pk_c_only}
  \begin{align}
    \xi_0\left(\mu,\eta,k\right)=&\max\left\{\mu-\eta\gamma_s\left(k\right)k^2,1/\left(\tilde{P}_m+1/\gamma_c\left(k\right)\right)\right\},\\
    \tilde{P}_c\left(k\right)=&\left\{\dfrac{1}{\xi_0\left(\mu_c^*,0,k\right)}-\dfrac{1}{\gamma_c\left(k\right)}\right\}^+,
  \end{align}
\end{subequations}

\begin{figure}[!t]
  \begin{algorithm}[H]
      \caption{Iterative Optimization Algorithm for Dual Variables in Communication-centric Scenarios}
      \label{alg:opt_dual_c}
      \begin{algorithmic}[1]
          \renewcommand{\algorithmicrequire}{\textbf{Input:}}
          \renewcommand{\algorithmicensure}{\textbf{Output:}}
          \REQUIRE Tolerance $\varepsilon_\mu,\varepsilon_\eta$ for $\mu,\eta$.
          \ENSURE Optimal $\eta_c^*$ and $\mu_c^*$.
          \STATE Initialize $\eta^{\left(0\right)}\leftarrow 0$, $\mu^{\left(0\right)}\leftarrow 0$, $j\leftarrow 0$.
          \WHILE {$\lvert\mu^{\left(j+1\right)}-\mu^{\left(j\right)}\rvert\geq\varepsilon_\mu$ or $\lvert\eta^{\left(j+1\right)}-\eta^{\left(j\right)}\rvert\geq\varepsilon_\eta$}
            \STATE $\mu_c\left(j\right)=\max\limits_{l}\ {\gamma_s\left(l\right)l^2\eta^{\left(j\right)}+\gamma_c\left(l\right)}$.
            \STATE Solve (\ref{xi_1_eq}) on $\left[\mu^{\left(j\right)},\mu_c\left(j\right)\right]$ by the bisection method to obtain $\mu^{\left(j+1\right)}$.
            \STATE $\eta_c\left(j\right)=\max\limits_{l}\ {\mu^{\left(j+1\right)}/\gamma_s\left(l\right)l^2}$.
            \STATE Solve (\ref{xi_2_eq}) on $\left[\eta^{\left(j\right)},\eta_c\left(j\right)\right]$ by the bisection method to obtain $\eta^{\left(j+1\right)}$ .
            \STATE $j\leftarrow j+1$.
          \ENDWHILE
          \STATE $\eta_c^*\leftarrow\eta^{\left(j\right)}$, $\mu_c^*\leftarrow\mu^{\left(j\right)}$.
      \end{algorithmic}
  \end{algorithm}
\end{figure}

{\noindent}where the optimal dual variable $\mu_c^*$ can be obtained by solving $\sum_{k=1}^{N/2-1}{\tilde{P}_c\left(k\right)}=1/2$. If the sensing constraint is still satisfied, i.e.,
\begin{equation}
  \sum_{k=1}^{\frac{N}{2}-1}{k^2\gamma_s\left(k\right)\tilde{P}_c\left(k\right)}\geq\frac{N\varsigma_0^2}{8\pi^2M{\Delta f}^2}:=\tilde{\varsigma}_0^2,
\end{equation}

{\noindent}then the optimal solution to (\ref{opt_Pk_c}) is $\tilde{P}_c^*\left(k\right)=\tilde{P}_c\left(k\right)$.

\subsubsection{Case B}The optimal Fisher information is considered, which requires more power allocated to subcarriers with higher normalized frequency, i.e., with larger $k^2\gamma_s\left(k\right)$. In this case, the index $k$ of the normalized frequency is first sorted as $k_l$ in ascending order, which satisfies
\begin{equation}
  \gamma_s\left(k_{l_1}\right)k_{l_1}^2\leq\gamma_s\left(k_{l_2}\right)k_{l_2}^2,\quad\forall l_1\leq l_2.
\end{equation}

{\noindent}Then, by defining $l_m=\lfloor N/2-1/\bigl(2\tilde{P}_m\bigr)\rfloor$ with $\lfloor\cdot\rfloor$ denoting the floor operator, the normalized power allocation for the $k_l$-th subcarrier is expressed as
\begin{equation}\label{opt_Pk_s_only}
  \tilde{P}_s\left(k_l\right)=
  \begin{cases}
    0,&1\leq k_l\leq k_{l_m}-1,\\
    \frac{1}{2}-\left(\frac{N}{2}-1-k_{l_m}\right)\tilde{P}_m,&k_l=k_{l_m},\\
    \tilde{P}_m,&k_{l_m}+1\leq k_l<\frac{N}{2}.
  \end{cases}
\end{equation}

If $\tilde{P}_s\left(k\right)$ cannot satisfy the sensing constraint in (P1-2), i.e.,
\begin{equation}
  \sum_{k=1}^{\frac{N}{2}-1}{k^2\gamma_s\left(k\right)\tilde{P}_s\left(k\right)}<\tilde{\varsigma}_0^2,
\end{equation}

{\noindent}then the problem is infeasible. Otherwise, the optimization is further conducted in \textit{Case C}.

\subsubsection{Case C}In this case, the sensing constraint cannot be satisfied by $\tilde{P}_c\left(k\right)$, while the problem is still feasible for $\tilde{P}_s\left(k\right)$. Consequently, a trade-off exist between communication and sensing metrics, and the optimal power allocation yielded by the KKT conditions is expressed as
\begin{equation}
  \tilde{P}_c^*\left(k\right)=\left\{\dfrac{1}{\xi_0\left(\mu_c^*,\eta_c^*,k\right)}-\dfrac{1}{\gamma_c\left(k\right)}\right\}^+.
\end{equation}

As the sensing constraint (\ref{opt_c_constr_sensing_2}) is always active in this case, the optimal dual variables $\mu_c^*$ and $\eta_c^*$ can be obtained by simultaneously solving
\begin{subequations}
  \begin{align}
    &\xi_1\left(\mu_c^*,\eta_c^*\right)=\frac{1}{2},\label{xi_1_eq}\\
    &\xi_2\left(\mu_c^*,\eta_c^*\right)=\tilde{\varsigma}_0^2,\label{xi_2_eq}
  \end{align}
\end{subequations}

{\noindent}where the functions $\xi_1\left(\mu,\eta\right)$ and $\xi_2\left(\mu,\eta\right)$ are defined as
\begin{subequations}
  \begin{align}
    \xi_1\left(\mu,\eta\right)=&\sum_{k=1}^{\frac{N}{2}-1}{\left\{\dfrac{1}{\xi_0\left(\mu,\eta,k\right)}-\dfrac{1}{\gamma_c\left(k\right)}\right\}^+},\label{xi_1_definition}\\
    \xi_2\left(\mu,\eta\right)=&\sum_{k=1}^{\frac{N}{2}-1}{k^2\gamma_s\left(k\right)\left\{\dfrac{1}{\xi_0\left(\mu,\eta,k\right)}-\dfrac{1}{\gamma_c\left(k\right)}\right\}^+}.\label{xi_2_definition}
  \end{align}
\end{subequations}

To obtain the optimal value of $\tilde{P}_c^*\left(k\right)$, an iterative optimization algorithm for dual variables is proposed, which solves (\ref{xi_1_eq}) and (\ref{xi_2_eq}) iteratively until the values of $\eta^{\left(j\right)}$ and $\mu^{\left(j\right)}$ converge. The algorithm is summarized in~\textbf{Algorithm~\ref{alg:opt_dual_c}}.


\subsection{Convergence Analysis}\label{converge_c}
While the monotonic convergence of~\textbf{Algorithm~\ref{alg:opt_alter_c}} cannot be derived theoretically, the convergence of~\textbf{Algorithm~\ref{alg:opt_dual_c}} is proven by the following \textit{Lemmas 1} and \textit{2}.

\begin{figure}[tp]
  \centering
  \includegraphics[width=0.48\textwidth]{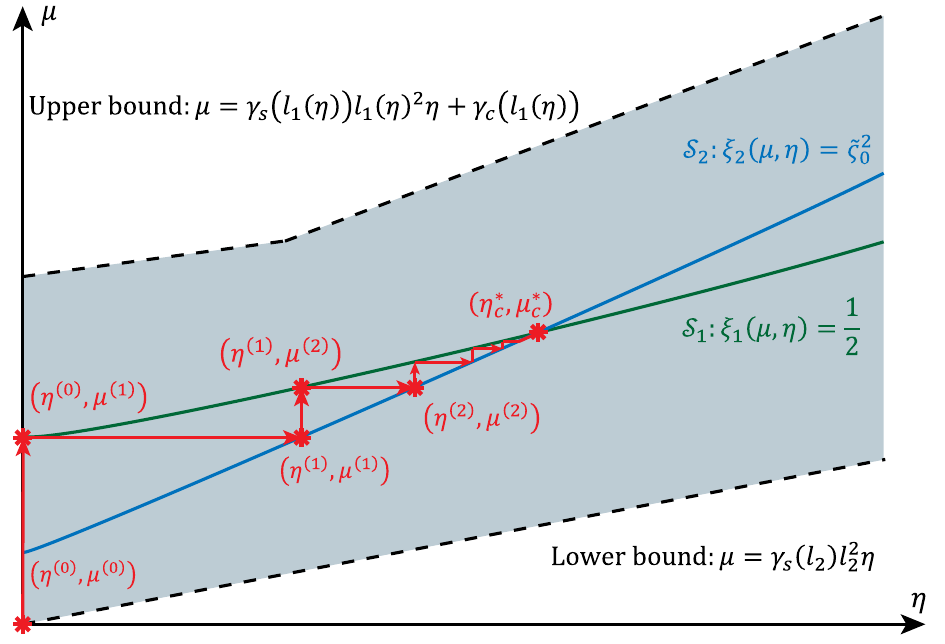}
  \caption{Sketch map of the iterative optimization algorithm for dual variables.}
  \label{fig:alter_opt}
\end{figure}

\textit{Lemma 1: }The optimal $\left(\mu_c^*,\eta_c^*\right)$ exists in the first quadrant, which is the intersection of curves
\begin{subequations}
  \begin{align}
    &\varGamma_1:\ \xi_1\left(\mu,\eta\right)=\frac{1}{2},\\
    &\varGamma_2:\ \xi_2\left(\mu,\eta\right)=\tilde{\varsigma}_0^2.
  \end{align}
\end{subequations}

\textit{Proof: }See Appendix~\ref{append_convergence_exist}.

\textit{Lemma 2: }The optimal $\left(\mu_c^*,\eta_c^*\right)$ lies in the region
\begin{equation}
  \gamma_s\left(l_2\right)l_2^2\eta\leq\mu\leq\gamma_s\left(l_1\left(\eta\right)\right)l_1\left(\eta\right)^2\eta+\gamma_c\left(l_1\left(\eta\right)\right),
\end{equation}

{\noindent}where the indices $l_1\left(\eta\right)$ and $l_2$ are obtained by
\begin{subequations}
  \begin{align}
    l_1\left(\eta\right)=&\mathop{\arg\ \max}\limits_{l}\ {\gamma_s\left(l\right)l^2\eta+\gamma_c\left(l\right)},\\
    l_2=&\mathop{\arg\ \min}\limits_{l}\ {\gamma_s\left(l\right)l^2}.
  \end{align}
\end{subequations}

\textit{Proof: }See Appendix~\ref{append_convergence_region}.

\textit{Proposition 1: }\textbf{Algorithm~\ref{alg:opt_dual_c}} converges to the optimal $\left(\mu_c^*,\eta_c^*\right)$.

\textit{Proof: }As illustrated in Fig.~\ref{fig:alter_opt}, \textit{Lemma 1} indicates the existence of $\left(\mu_c^*,\eta_c^*\right)$, and \textit{Lemma 2} guarantees that $\left(\mu^{\left(i\right)},\eta^{\left(i\right)}\right)$ will not leave the feasible region during the iteration. Furthermore, we have the initial points $\mu^{\left(0\right)}<\mu_c^*$, $\eta^{\left(0\right)}<\eta_c^*$ since $\mu_c^*>0$, $\eta_c^*>0$. Supposing that $\eta^{\left(j\right)}<\eta_c^*$, solving $\xi_1\left(\mu^{\left(j+1\right)},\eta^{\left(j\right)}\right)=1/2$ yields $\mu^{\left(j\right)}<\mu^{\left(j+1\right)}<\mu_c^*$. Similarly, assuming that $\mu^{\left(j+1\right)}<\mu_c^*$, solving $\xi_2\left(\mu^{\left(j+1\right)},\eta^{\left(j+1\right)}\right)=\tilde{\varsigma}_0^2$ yields $\eta^{\left(j\right)}<\eta^{\left(j+1\right)}<\eta_c^*$. Therefore, the distance between the current solution and the optimal solution declines as the algorithm proceeds, i.e.,
\begin{subequations}
  \begin{align}
    &\lvert \mu^{\left(j+1\right)}-\mu_c^* \rvert < \lvert \mu^{\left(j\right)}-\mu_c^* \rvert,\\
    &\lvert \eta^{\left(j+1\right)}-\eta_c^* \rvert < \lvert \eta^{\left(j\right)}-\eta_c^* \rvert.
  \end{align}
\end{subequations}

In consequence, the convergence of the algorithm is proven by the Monotone and Boundedness theorem~\cite{hughes_calculus}.


\subsection{Computational Complexity Analysis}\label{complexity_c}
The computational complexity of solving (P1) mainly comprises that of the golden search for (P1-1) and the iterative optimization for (P1-2) in each iteration. On the one hand, since the error of the golden search declines geometrically, the complexity of solving (P1-1) is $\mathcal{O}\left(\log\left(1/\varepsilon_b\right)\right)$ for a desired error threshold of $\varepsilon_b$. On the other hand, the complexity of bisection method in each iteration is $\mathcal{O}\left(\log\left(1/\varepsilon_\mu\right)+\log\left(1/\varepsilon_\eta\right)\right)$. Besides, numerical simulations indicate that the residual between $\left(\mu^{\left(i\right)},\eta^{\left(i\right)}\right)$ and $\left(\mu_c^*,\eta_c^*\right)$ declines almost exponentially. Therefore, the computational complexity of solving (P1-2) is $\mathcal{O}\bigl(\left(\log\left(1/\varepsilon_\mu\right)+\log\left(1/\varepsilon_\eta\right)\right)^2\bigr)$. In summary, supposing that~\textbf{Algorithm~\ref{alg:opt_alter_c}} converges after $K$ iterations, the overall computational complexity of solving (P1) is given by $\mathcal{O}\bigl(K\bigl(\log\left(1/\varepsilon_b\right)+\left(\log\left(1/\varepsilon_\mu\right)+\log\left(1/\varepsilon_\eta\right)\right)^2\bigr)\bigr)$.

\begin{figure}[!t]
  \begin{algorithm}[H]
      \caption{BCD Algorithm for Sensing-centric Scenario}
      \label{alg:opt_alter_s}
      \begin{algorithmic}[1]
          \renewcommand{\algorithmicrequire}{\textbf{Input:}}
          \renewcommand{\algorithmicensure}{\textbf{Output:}}
          \REQUIRE Tolerance $\epsilon_I$. Initial solution $b^{\left(0\right)}$ and $\tilde{P}^{\left(0\right)}\left(k\right)$.
          \ENSURE Optimal $b_s^*$ and $\tilde{P}_s^*\left(k\right)$ to~(P2).
          \STATE $i\leftarrow 0$, $I^{\left(0\right)}\leftarrow 0$.
          \WHILE{$\lvert I^{\left(i+1\right)}-I^{\left(i\right)} \rvert\geq\epsilon_I$}
            \STATE Given $\tilde{P}^{\left(i\right)}\left(k\right)$, solve (P2-1) to obtain $b^{\left(i+1\right)}$.
            \STATE Given $b^{\left(i+1\right)}$, solve (P2-2) to obtain $\tilde{P}^{\left(i+1\right)}\left(k\right)$ and $I^{\left(i+1\right)}$.
            \STATE $i\leftarrow i+1$.
          \ENDWHILE
          \STATE $b_c^*\leftarrow b^{\left(i\right)}$, $\tilde{P}_c^*\left(k\right)\leftarrow \tilde{P}^{\left(i\right)}\left(k\right)$.
      \end{algorithmic}
  \end{algorithm}
\end{figure}

\section{Sensing-centric Resource Allocation}\label{power_allocation_s}
In this section, we investigate the sensing-centric power allocation problem, which is first formulated in Subsection~\ref{power_allocation_s_joint}. Then, the joint power allocation problem is decomposed into a sub-problem for DC bias and a sub-problem for power allocation on subcarriers, which are elaborated in Subsections~\ref{power_allocation_s_bias} and~\ref{power_allocation_s_subcarrier}, respectively.

\subsection{Joint Power Allocation Problem}\label{power_allocation_s_joint}
For the sensing-centric scenario, the joint power allocation problem achieves the optimal sensing performance and guarantees the communication performance simultaneously, which is formulated as
\begin{subequations}\label{opt_s}
  \begin{align}
    &\text{(P2):}& &\mathop{\max}\limits_{b,\tilde{P}\left(k\right)}\ & &I\left(b,\tilde{P}\left(k\right)\right),&\label{opt_s_obj}\\
    & & &\text{s.t.}\ & &C\left(b,\tilde{P}\left(k\right)\right)\geq C_0,&\label{opt_s_constr_comm}\\
    & & &\ & &0\leq b\leq\sqrt{P},&\label{opt_s_constr_bias}\\
    & & &\ & &0\leq\tilde{P}\left(k\right)\leq \tilde{P}_m,\label{opt_s_constr_power_each}&\\
    & & &\ & &\sum_{k=1}^{\frac{N}{2}-1}{\tilde{P}\left(k\right)}=\frac{1}{2},&\label{opt_s_constr_power_total}
  \end{align}
\end{subequations}

{\noindent}where $C_0$ is the desired spectral efficiency. Constraints (\ref{opt_s_constr_comm}), (\ref{opt_s_constr_bias}), (\ref{opt_s_constr_power_each}), and (\ref{opt_s_constr_power_total}) correspond to the communication performance, the DC bias, the power allocation for subcarriers, and the total power, respectively.

Similar to the communication-centric scenario, the joint optimization for $b$ and $\tilde{P}\left(k\right)$ is also non-convex in~(P2). Therefore, we also decompose (P2) into a sub-problem for $b$ and a sub-problem for $\tilde{P}\left(k\right)$, and then adopt a BCD algorithm to solve (P2), which is summarized in~\textbf{Algorithm~\ref{alg:opt_alter_s}}.

\subsection{Power Allocation for DC Bias}\label{power_allocation_s_bias}
The sub-problem for DC bias is formulated as
\begin{subequations}\label{opt_b_s}
  \begin{align}
    &\text{(P2-1):}& &\mathop{\max}\limits_{b}\ & &I\left(b,\tilde{P}\left(k\right)\right),&\\
    & & &\text{s.t.}\ & &0\leq b\leq\sqrt{P}.&
  \end{align}
\end{subequations}

Similarly, the approximation of~(\ref{gamma_c}) can be readily applied to~(\ref{opt_b_s}). Therefore, by golden search through the feasible interval, the optimal $b_s^*$ for the sensing-centric scenario can be obtained.

\subsection{Power Allocation for Subcarriers}\label{power_allocation_s_subcarrier}
Once $b_s^*$ is obtained by solving (P2-1), the normalized power allocation for subcarriers is further optimized to maximize the Fisher information. Similarly, the clipping noise PSD $P_{w_p}\left(k\right)$ is still assumed to be a constant for the sake of concise results. In consequence, the sub-problem to optimize $\tilde{P}\left(k\right)$ is formulated as
\begin{subequations}\label{opt_Pk_s}
  \begin{align}
    &\text{(P2-2):}& &\mathop{\max}\limits_{\tilde{P}\left(k\right)}& &I\left(b,\tilde{P}\left(k\right)\right),&\\
    & & &\text{s.t.}\ & &C\left(b,\tilde{P}\left(k\right)\right)\geq C_0,&\label{opt_s_constr_comm_2}\\
    & & &\ & &\sum_{k=1}^{\frac{N}{2}-1}{\tilde{P}\left(k\right)}=\frac{1}{2},&\\
    & & &\ & &0\leq\tilde{P}\left(k\right)\leq \tilde{P}_m.&
  \end{align}
\end{subequations}

As discussed in Section~\ref{power_allocation_c_subcarrier}, $C\bigl(b,\tilde{P}\left(k\right)\bigr)$ is a concave function of $\tilde{P}\left(k\right)$, which means the feasible region given by~(\ref{opt_s_constr_comm_2}) is a convex set. In addition, the objective and power constraints form a linear programming problem, and (\ref{opt_Pk_s}) is therefore a convex optimization problem. Similar to its communication-centric counterpart, three cases derived by the KKT conditions are considered as follows.

\subsubsection{Case D} The communication constraint is first assumed to be inactive, and (\ref{opt_Pk_s}) is reduced to a linear programming problem, whose optimal solution is given by (\ref{opt_Pk_s_only}). If the communication constraint is satisfied by $\tilde{P}_s\left(k\right)$, i.e.,
\begin{equation}
  \sum_{k=1}^{\frac{N}{2}-1}{\log\left(1+\gamma_c\left(k\right)\tilde{P}_s\left(k\right)\right)}\geq C_0BT_o:=\tilde{C}_0,
\end{equation}

{\noindent}then the optimal solution is $\tilde{P}_s^*\left(k\right)=\tilde{P}_s\left(k\right)$.

\begin{figure}[!t]
  \begin{algorithm}[H]
      \caption{Iterative Optimization Algorithm for Dual Variables in Sensing-centric Scenarios}
      \label{alg:opt_dual_s}
      \begin{algorithmic}[1]
          \renewcommand{\algorithmicrequire}{\textbf{Input:}}
          \renewcommand{\algorithmicensure}{\textbf{Output:}}
          \REQUIRE Tolerance $\varepsilon_\mu,\varepsilon_\eta$ for $\mu,\eta$.
          \ENSURE Optimal $\eta_s^*$ and $\mu_s^*$.
          \STATE Initialize $\eta^{\left(0\right)}\leftarrow 0$, $\mu^{\left(0\right)}\leftarrow 0$, $j\leftarrow 0$.
          \WHILE{$\lvert\mu^{\left(j+1\right)}-\mu^{\left(j\right)}\rvert\geq\varepsilon_\mu$ or $\lvert\eta^{\left(j+1\right)}-\eta^{\left(j\right)}\rvert\geq\varepsilon_\eta$}
            \STATE $\mu_s\left(j\right)=\max\limits_{l}\ {\gamma_s\left(l\right)l^2+\gamma_c\left(l\right)\eta^{\left(j\right)}}$.
            \STATE Solve (\ref{psi_1_eq}) on $\left[\mu^{\left(j\right)},\mu_s\left(j\right)\right]$ by the bisection method to obtain $\mu^{\left(j+1\right)}$.
            \STATE $\eta_s\left(j\right)=\max\limits_{l}\ {\left(\tilde{P}_m+1/\gamma_c\left(l\right)\right)\left(\mu^{\left(j+1\right)}-\gamma_s\left(l\right)l^2\right)}$.
            \STATE Solve (\ref{psi_2_eq}) on $\left[\eta^{\left(j\right)},\eta_s\left(j\right)\right]$ by the bisection method to obtain $\eta^{\left(j+1\right)}$.
            \STATE $j\leftarrow j+1$.
          \ENDWHILE
          \STATE $\eta_s^*\leftarrow\eta^{\left(j\right)}$, $\mu_s^*\leftarrow\mu^{\left(j\right)}$.
      \end{algorithmic}
  \end{algorithm}
\end{figure}

\subsubsection{Case E} The optimal spectral efficiency is considered and the optimal power allocation is obtained by the conventional water-filling method, as shown in~(\ref{opt_Pk_c_only}). If $\tilde{P}_c\left(k\right)$ cannot satisfy the communication constraint, i.e.,
\begin{equation}
  \sum_{k=1}^{\frac{N}{2}-1}{\log\left(1+\gamma_c\left(k\right)\tilde{P}_c\left(k\right)\right)}<\tilde{C}_0,
\end{equation}

{\noindent}then the problem is infeasible. Otherwise, the optimization is further conducted in \textit{Case F}.

\subsubsection{Case F}In this case, the communication constraint cannot be satisfied by $\tilde{P}_s\left(k\right)$, while the problem is still feasible for $\tilde{P}_c\left(k\right)$. In consequence, the optimal power allocation yielded by the KKT conditions is expressed as
\begin{subequations}
  \begin{align}
    \psi_0\left(\mu,\eta,k\right)=&\max\left\{\left(\mu-\gamma_s\left(k\right)k^2\right)/\eta,\right.\notag\\
    &\left.1/\left(\tilde{P}_m+1/\gamma_c\left(k\right)\right)\right\},\\
    \tilde{P}_s^*\left(k\right)=&\left\{\dfrac{1}{\psi_0\left(\mu_s^*,\eta_s^*,k\right)}-\dfrac{1}{\gamma_c\left(k\right)}\right\}^+.
  \end{align}
\end{subequations}

As the communication constraint is always active in this case, the optimal dual variables $\mu_s^*$ and $\eta_s^*$ can be obtained by simultaneously solving
\begin{subequations}
  \begin{align}
    &\sum_{k=1}^{\frac{N}{2}-1}{\left\{\dfrac{1}{\psi_0\left(\mu_s^*,\eta_s^*,k\right)}-\dfrac{1}{\gamma_c\left(k\right)}\right\}^+}=\frac{1}{2},\label{psi_1_eq}\\
    &\sum_{k=1}^{\frac{N}{2}-1}{\log\left(1+\gamma_c\left(k\right)\left\{\dfrac{1}{\psi_0\left(\mu_s^*,\eta_s^*,k\right)}-\dfrac{1}{\gamma_c\left(k\right)}\right\}^+\right)}=\tilde{C}_0.\label{psi_2_eq}
  \end{align}
\end{subequations}

The value of $\tilde{P}_s^*\left(k\right)$ can also be obtained by an iterative optimization algorithm, which is summarized in \textbf{Algorithm~\ref{alg:opt_dual_s}}. Due to space limitations, the proof of convergence and complexity analysis are omitted here, which resembles that of its communication-centric counterpart.

\begin{table}[tp]
  \centering
  \renewcommand\arraystretch{1.2}
  \caption{Simulation Configurations}
  \begin{tabular}{c|c|c}
      \hline \hline
      Parameter & Notation & Value \\
      \hline
      OFDM symbols in each frame & $M$ & 64 \\
      Subcarriers in each OFDM symbol & $N$ & 1024 \\
      Subcarrier spacing & $\Delta f$ & 0.2 MHz \\
      Duration of a guard interval & $T_g$ & 2 \textmu s \\
      Duration of an OFDM symbol & $T_o$ & 7 \textmu s \\
      OFDM bandwidth & $B$ & 204.8 MHz \\
      Total power & $P$ & 1 W \\
      \hline
      Target distance & $L$ & 200 m \\
      Optical wavelength & $\lambda$ & 905 nm \\
      Atmospheric attenuation & $\alpha V^{\beta}$ & -12.8 dB/km \\
      Refractive index & $C_n^2$ & $5\times 10^{-14}\ \text{m}^{-2/3}$ \\
      Beam divergence angle & $\theta$ & 0.5 mrad \\
      Receiver aperture & $A$ & $10\ \text{cm}^2$ \\
      Target reflectivity & $\mathfrak{R}$ & 0.5 \\
      Transmitter gain & $G_T$ & 1 \\
      Receiver gain & $G_R$ & 10 \\
      \hline \hline
  \end{tabular}
  \label{tab:sim_config}
\end{table}


\section{Numerical Results}\label{numerical_simulation}
\begin{figure}[tp]
  \centering
  \includegraphics[width=0.48\textwidth]{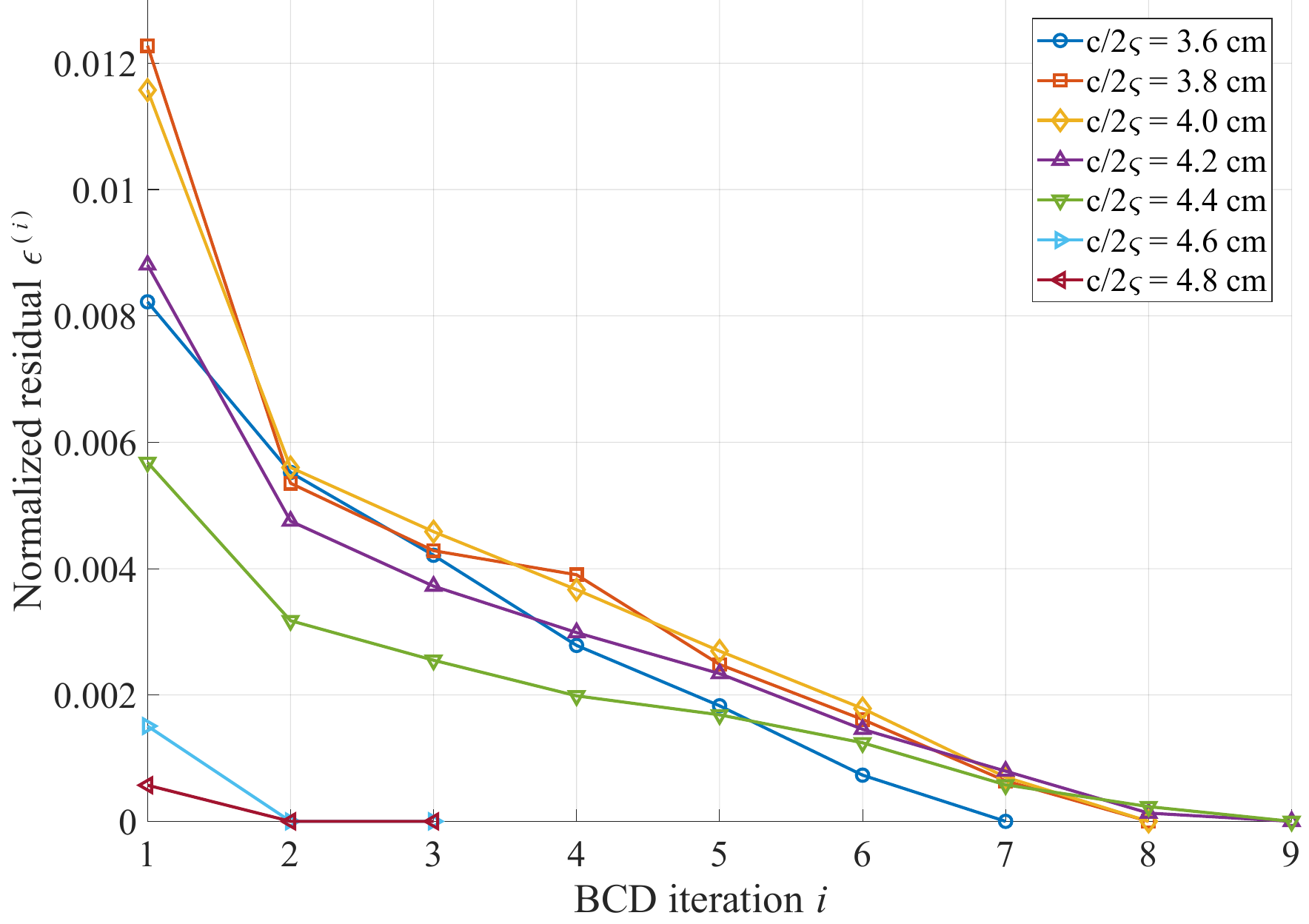}
  \caption{Convergence of the spectral efficiency in the communication-centric scenario.}
  \label{fig:converge_opt_c}
\end{figure}

This section provides numerical results to evaluate the proposed DCO-OFDM scheme for FSO-ISAC, and Table~\ref{tab:sim_config} shows configurations for system parameters. We consider a terrestrial scenario where a 905 nm LiDAR is adopted as the FSO-ISAC transceiver. A guard interval of $T_g$ is concatenated in front of each OFDM symbol and guarantees an unambiguous distance of larger than $300\ \text{m}$. Moreover, due to the narrow and directional characteristics of laser beams adopted by FSO, we only consider line-of-sight (LoS) channels, i.e., $\tilde{H}_c\left(k\right)=\tilde{H}_s\left(k\right)=1$. Supposing that the target is located at a distance of $L=200\ \text{m}$, the stationary channel gains can be calculated as -2.2 dB and -23.2 dB for the communication subsystem and the sensing subsystem, respectively. Subsequently, we first demonstrate the convergence of the proposed algorithms in Subsection~\ref{sim_converge} and then characterize the optimal resource allocation in Subsection~\ref{sim_opt}.

\subsection{Convergence Results}\label{sim_converge}

\begin{figure}[tp]
  \centering
  \includegraphics[width=0.48\textwidth]{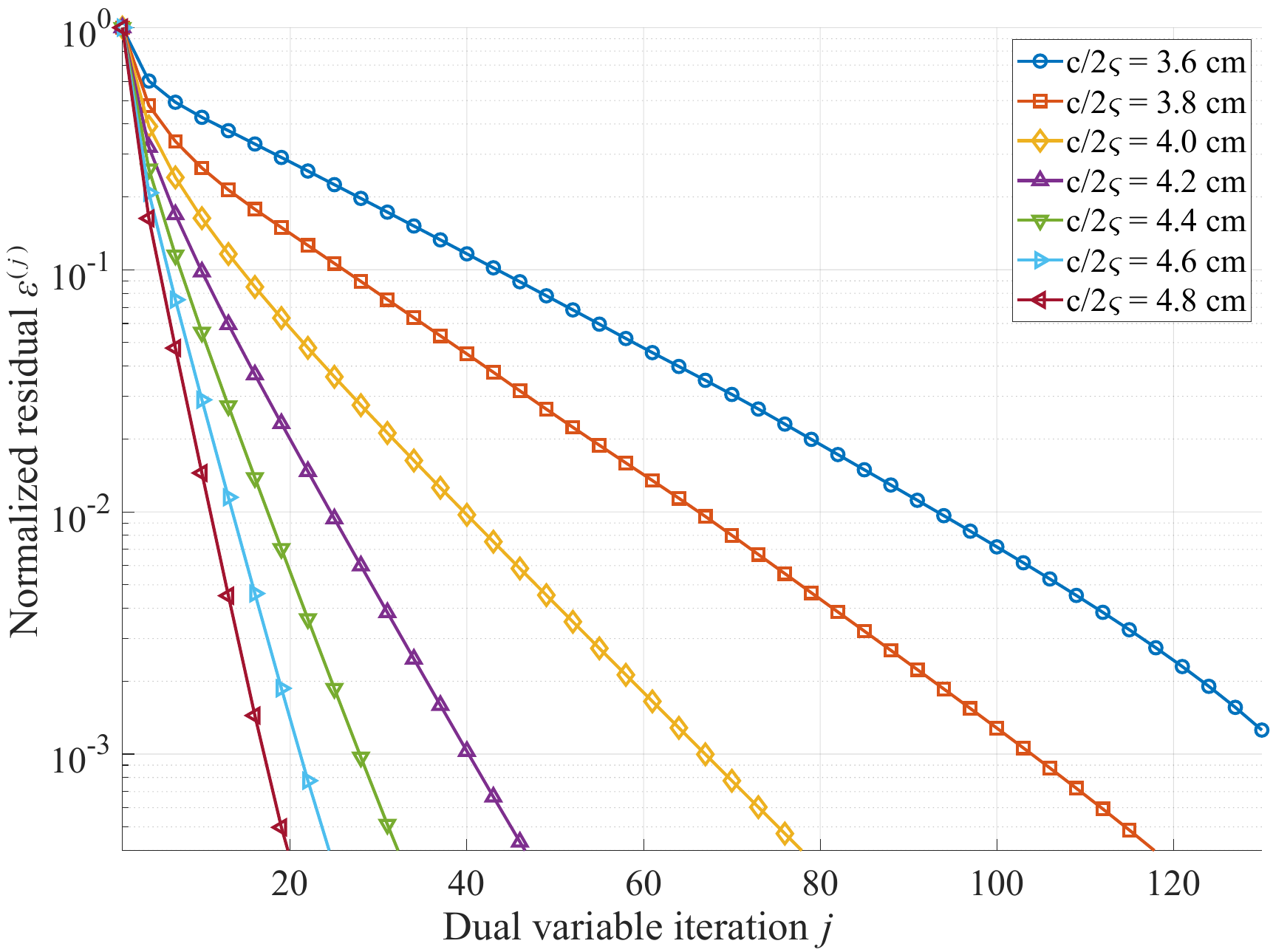}
  \caption{Convergence of dual variables in the communication-centric scenario.}
  \label{fig:converge_opt_dual_c}
\end{figure}

Fig.~\ref{fig:converge_opt_c} illustrates the convergence of the spectral efficiency $C^{\left(i\right)}$ in the communication-centric scenario, which demonstrates the convergence of the proposed BCD algorithm, i.e., \textbf{Algorithm~\ref{alg:opt_alter_c}}. First, a normalized residual for $C^{\left(i\right)}$ is defined as
\begin{equation}
  \epsilon^{\left(i\right)}=\frac{\lvert C^{\left(i\right)}-C^*\rvert}{C^*},
\end{equation}

{\noindent}which depicts the distance between the current spectral efficiency and the optimal spectral efficiency. For a desired precision ranging from 3.6 cm to 4.8 cm, the spectral efficiency $C^{\left(i\right)}$ can converge within 10 iterations, and then the optimal power allocation can be achieved for the joint optimization of $b$ and $\tilde{P}\left(k\right)$.

\begin{figure}[tp]
  \centering
  \includegraphics[width=0.48\textwidth]{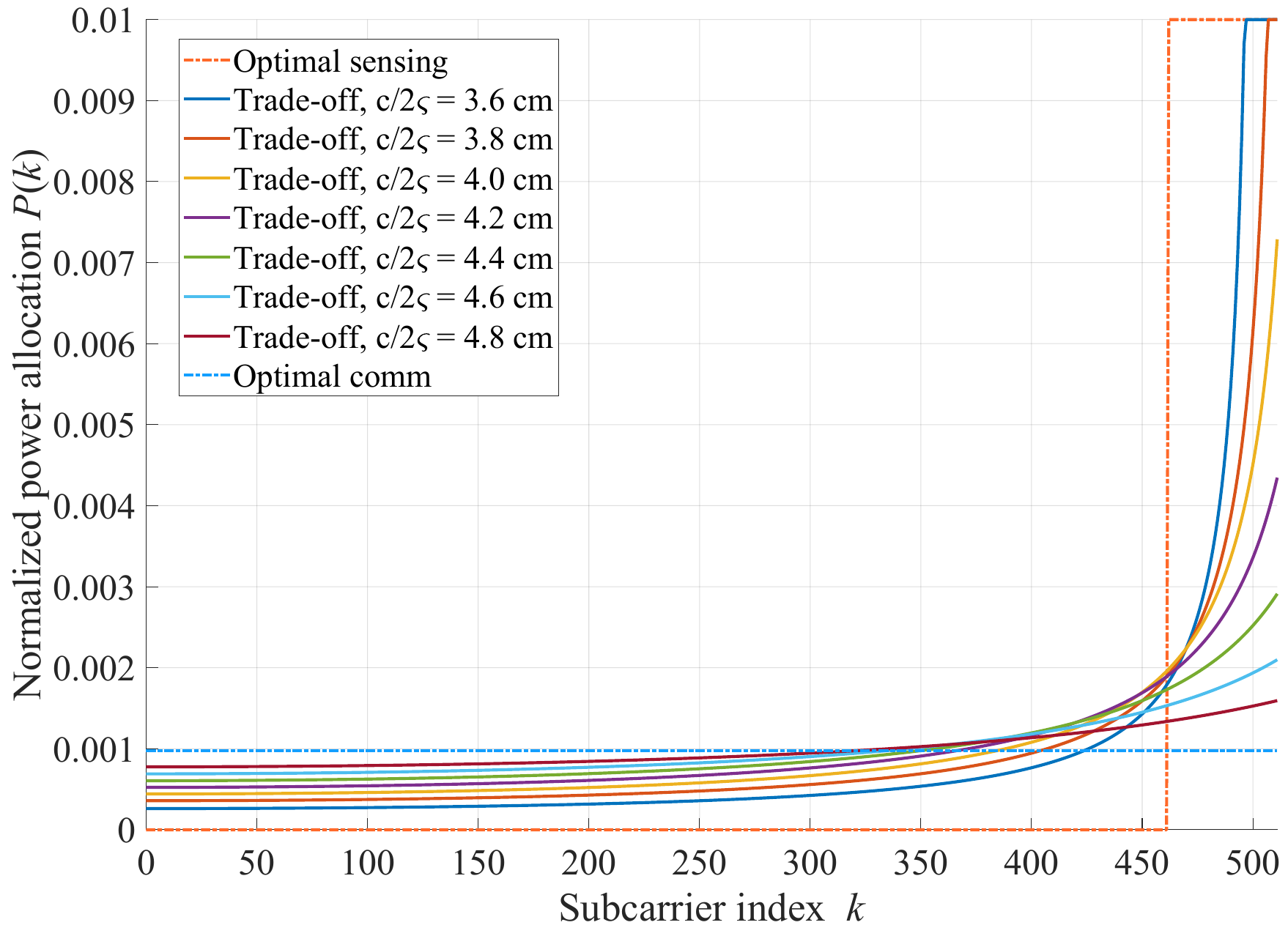}
  \caption{Power allocation on subcarriers for the communication-centric scenario.}
  \label{fig:power_allocation_c}
\end{figure}

Fig.~\ref{fig:converge_opt_dual_c} illustrates the convergence of dual variables in \textit{Case C} of the communication-centric scenario, which demonstrates the convergence of~\textbf{Algorithm~\ref{alg:opt_dual_c}}. To describe the speed of convergence, we define a normalized residual as 
\begin{equation}
  \varepsilon^{\left(j\right)}=\frac{\lvert \mu^{\left(j\right)}-\mu_c^* \rvert}{\lvert \mu^{\left(0\right)}-\mu_c^* \rvert}+\frac{\lvert \eta^{\left(j\right)}-\eta_c^* \rvert}{\lvert \eta^{\left(0\right)}-\eta_c^* \rvert},
\end{equation}

{\noindent}which depicts the distance between current dual variables and optimal dual variables. As shown in Fig.~\ref{fig:converge_opt_dual_c}, the residual declines almost geometrically as the iteration proceeds. In addition, when the desired sensing precision becomes higher, the distance between $\left(\mu^*,\eta^*\right)$ and $\left(\mu^{\left(0\right)},\eta^{\left(0\right)}\right)$ also increases. In consequence, the convergence of~\textbf{Algorithm~\ref{alg:opt_alter_c}} also slows down, as \textbf{Algorithm~\ref{alg:opt_dual_c}} is an essential part of \textbf{Algorithm~\ref{alg:opt_alter_c}} considering the computational complexity.

\subsection{Optimal Resource Allocation}\label{sim_opt}

\begin{figure}[tp]
  \centering
  \includegraphics[width=0.48\textwidth]{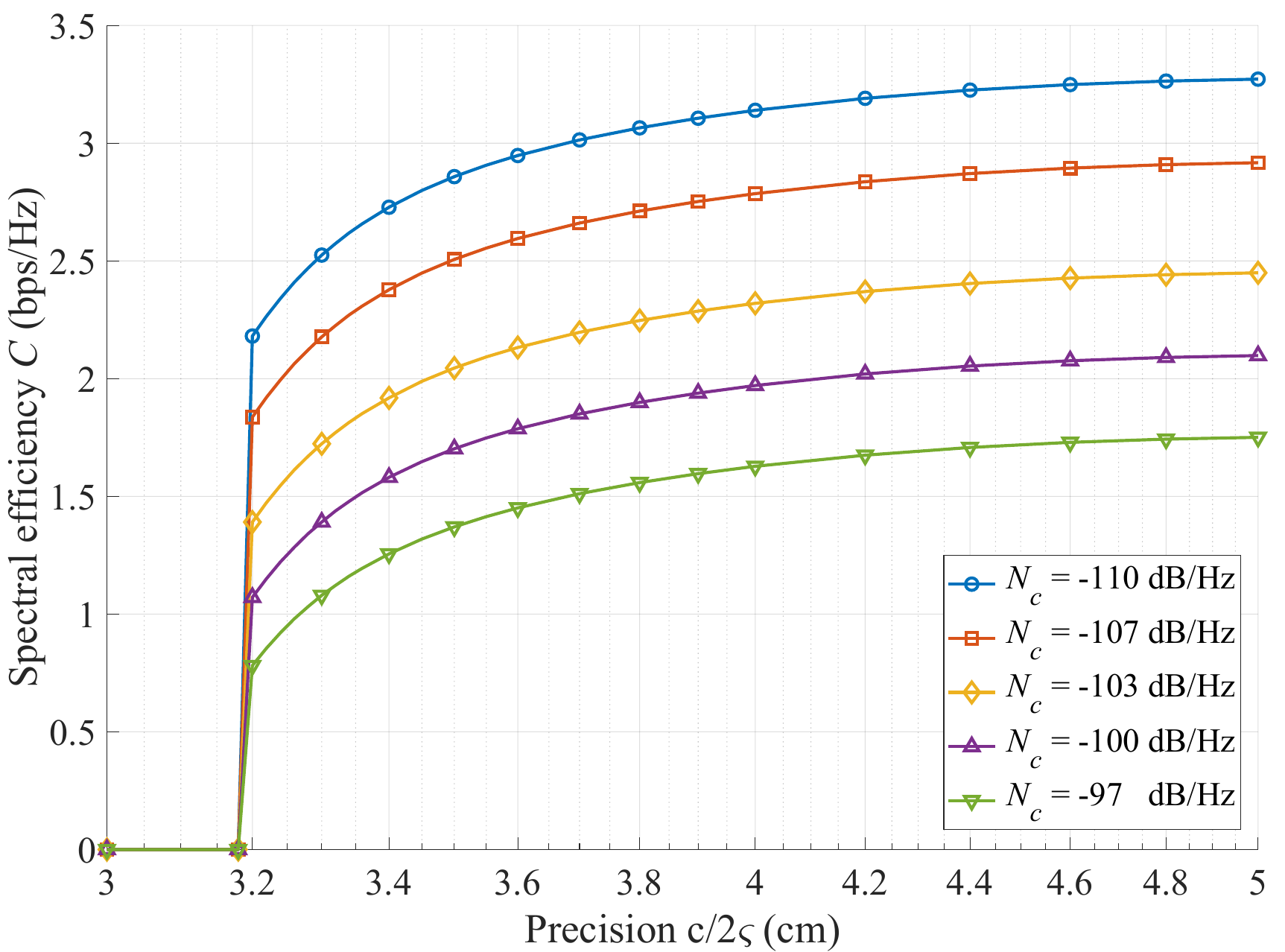}
  \caption{Spectral efficiency for the communication-centric scenario with respect to different sensing precisions.}
  \label{fig:spectral_efficiency}
\end{figure}

Fig.~\ref{fig:power_allocation_c} shows the normalized power allocation for subcarriers in the communication-centric scenario with respect to different sensing precisions, where system parameters are set to $N_c=N_s=-100\ \text{dB/Hz}$ and $\tilde{P}_m=0.01$. Due to the Hermitian symmetry in the frequency domain, the subcarrier index $k$ is restricted to $1\leq k\leq N/2-1$. Meanwhile, the optimal power allocation schemes for sole communication or sensing are also illustrated for a comparison. As described in~(\ref{opt_Pk_c_only}), the water-filling method yields a nearly uniform distribution for the communication-centric power allocation, while more power is allocated to subcarriers with higher normalized frequency to obtain a superior sensing precision. Moreover, the curvature of power allocation curves increases continuously as the desired sensing precision becomes higher, which embodies the compromise between communication and sensing functionalities.

\begin{figure}[tp]
  \centering
  \includegraphics[width=0.48\textwidth]{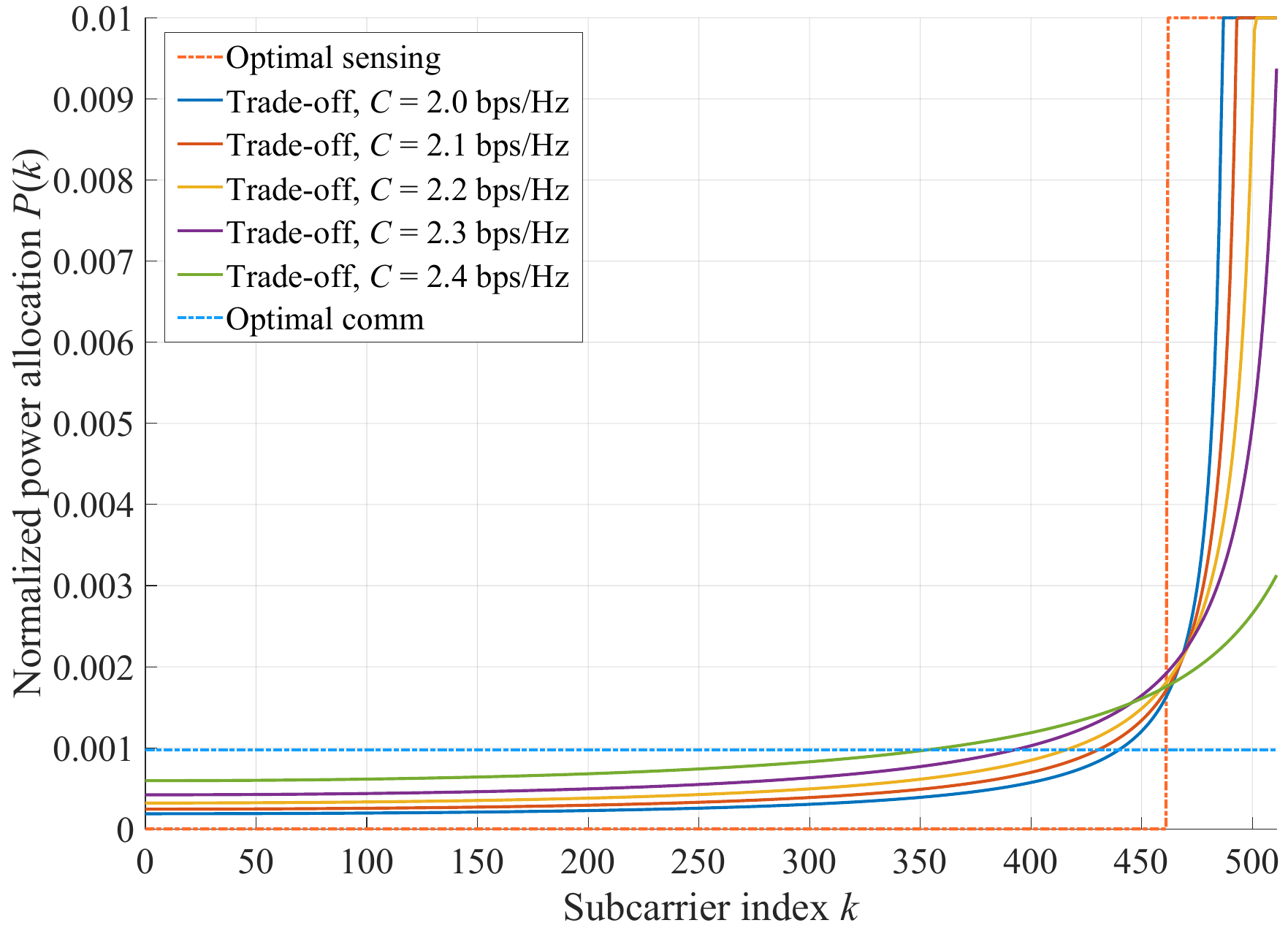}
  \caption{Power allocation on subcarriers for the sensing-centric scenario.}
  \label{fig:power_allocation_s}
\end{figure}

Fig.~\ref{fig:spectral_efficiency} illustrates the optimal spectral efficiency in the communication-centric scenario with respect to the desired sensing precision. For a desired sensing precision of smaller than $3.2\ \text{cm}$, the sensing constraint (\ref{opt_c_constr_sensing_2}) cannot be satisfied, and we set the spectral efficiency as zero. In addition, for $3.2\ \text{cm}\leq c/2\varsigma\leq 4.2\ \text{cm}$, the spectral efficiency grows rapidly as the feasible region given by (\ref{opt_c_constr_sensing_2}) is enlarged. However, for a desired sensing precision of larger than $4.2\ \text{cm}$, the spectral efficiency nearly reaches saturation, which cannot be further improved by increasing $c/2\varsigma$. Furthermore, even for the optimal sensing performance in \textit{Case B}, the FSO-ISAC scheme can still provide communication capability, and therefore a breaking point exists near $c/2\varsigma=3.2\ \text{cm}$ on each curve.

\begin{figure}[tp]
  \centering
  \includegraphics[width=0.48\textwidth]{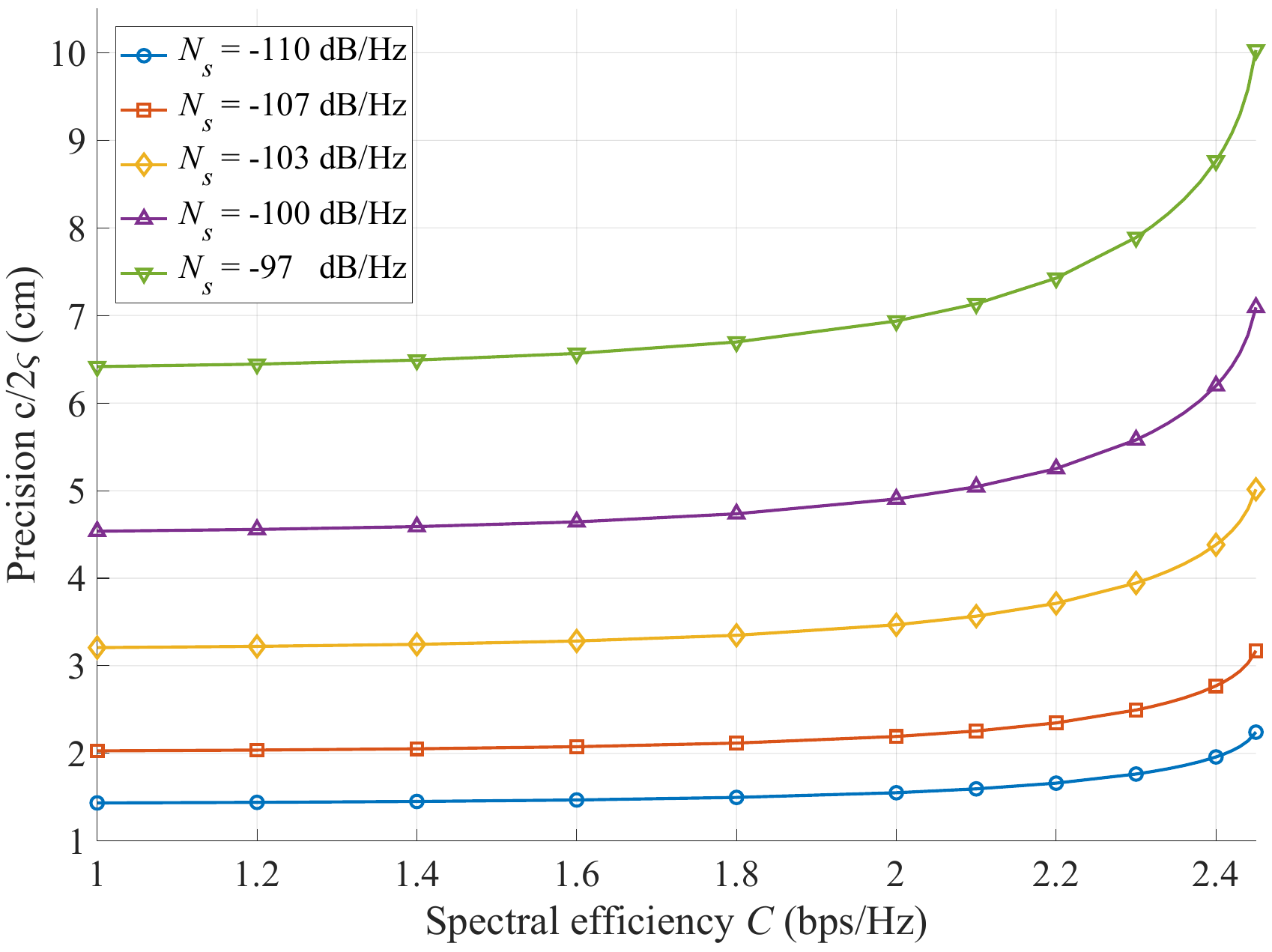}
  \caption{Sensing precision for the sensing-centric scenario with respect to different spectral efficiencies.}
  \label{fig:precision}
\end{figure}

Fig.~\ref{fig:power_allocation_s} shows the normalized power allocation for subcarriers in the sensing-centric scenario with respect to different spectral efficiencies, where system parameters are also set to $N_c=N_s=-100\ \text{dB/Hz}$ and $\tilde{P}_m=0.01$. Similar to the communication-centric scenario, the curvature of power allocation curves decreases continuously as the desired spectral efficiency increases, and a flexible communication performance can be obtained by tuning $C_0$. However, only slight differences exist between the optimal sensing scenario and the trade-off scenario for $C\leq 2.0\ \text{bps/Hz}$, while the curvature changes sharply for $C\geq 2.3\ \text{bps/Hz}$, which demands delicate tuning techniques for parameter setup.

Fig.~\ref{fig:precision} illustrates the optimal sensing precision in the sensing-centric scenario with respect to different desired spectral efficiencies. For a desired spectral efficiency of less than $1.8\ \text{bps/Hz}$, the optimal sensing precision is marginal, and decreasing the desired spectral efficiency does not affect the sensing precision significantly. However, for a desired spectral efficiency of larger than $2.0\ \text{bps/Hz}$, the sensing performance deteriorates dramatically as the desired spectral efficiency increases. Nonetheless, even for the optimal communication performance in \textit{Case E}, the FSO-ISAC scheme can still provide sensing ability, which yields the bounded sensing precisions in Fig.~\ref{fig:precision}.

In summary, numerical results demonstrate the proposed algorithms and reveal the trade-off between communication and sensing functionalities. Moreover, the communication-centric problem (P1) and the sensing-centric problem (P2) can be regarded as a dual to each other, which brings convenience for the system parameter setup. By tuning the desired sensing precision or spectral efficiency, an individual curve can be obtained in both Fig.~\ref{fig:power_allocation_c} and Fig.~\ref{fig:power_allocation_s}, whose communication and sensing performance metrics are completely the same. Therefore, based on the flexible algorithms proposed in this paper, the DCO-OFDM scheme provides an efficient approach to FSO-ISAC, considering different requirements of communication and sensing subsystems.


\section{Conclusion}\label{conclusion}
In this paper, a DCO-OFDM scheme was proposed for FSO-ISAC, whose performance metrics were derived for communication and sensing, i.e., the spectral efficiency and the Fisher information. The clipping noise of DCO-OFDM was modelled as additive colored Gaussian noise to obtain the expressions of SNR in these performance metrics. Additionally, joint power allocation problems were formulated in both communication-centric and sensing-centric scenarios, which were decomposed and solved by BCD algorithms. Moreover, case-by-case solutions to these problems were provided, and iterative algorithms were proposed to attain the optimal dual variables in the closed-form expressions of power allocation. Furthermore, numerical simulations demonstrated the proposed algorithms and revealed the trade-off between communication and sensing functionalities. Consequently, the proposed DCO-OFDM scheme could meet different requirements of communication and sensing subsystems, and therefore provide a flexible and efficient approach to FSO-ISAC.


{\appendices
\section{Proof of \textbf{Theorem 1}}\label{append_crb}

For simplicity, the signal is vectorized for the following derivation in this appendix. Denoting the signal without noise as $\bm{s}\left(\tau_0\right)$, the received sensing signal can be written as
\begin{equation}
  \bm{y}=\bm{s}\left(\tau_0\right)+\bm{w},
\end{equation}

{\noindent}where $\bm{w}$ denotes additive colored Gaussian noise, i.e., $\bm{w}\sim\mathcal{N}\left(\bm{0},\bm{C}\right)$, and the parameter $\tau_0$ is the ToF to be estimated. Then, the Fisher information is given by~\cite{kay_Fundamentals}
\begin{equation}\label{Fisher_info_orig}
  I=-\mathbb{E}_{\bm{y}}\left(\frac{\partial^2\log p\left(\bm{y};\tau_0\right)}{\partial\tau_0^2}\right)=\frac{\partial\bm{s}\left(\tau_0\right)^{\mathcal{H}}}{\partial\tau_0}\bm{C}^{-1}\frac{\partial\bm{s}\left(\tau_0\right)}{\partial\tau_0}.
\end{equation}

Denoting $\bm{W}$ as the IDFT matrix, whose element in the $k$-th row and the $n$-th column is $\left(\bm{W}\right)_{k,n}=\exp\left(j2\pi nk/N\right)/\sqrt{N}$, the covariance matrix of $\bm{w}$ can be expressed in a frequency-domain form as
\begin{equation}
  \bm{C}=\bm{W}\text{diag}\left(\bm{S}_{\bm{w}}\right)\bm{W}^{\mathcal{H}},
\end{equation}

{\noindent}where ${\bm{S}}_{\bm{w}}$ is the PSD vector of $\bm{w}$, and $\text{diag}\left({\bm{S}}_{\bm{w}}\right)$ denotes a diagonal matrix whose diagonal elements are ${\bm{S}}_{\bm{w}}$.

Substituting $\bm{C}$ in~(\ref{Fisher_info_orig}) further yields
\begin{equation}
  \begin{split}
    I&=\left(\bm{W}^{\mathcal{H}}\frac{\partial\bm{s}\left(\tau_0\right)}{\partial\tau_0}\right)^{\mathcal{H}}\text{diag}\left(\bm{S}_{\bm{w}}\right)^{-1}\left(\bm{W}^{\mathcal{H}}\frac{\partial\bm{s}\left(\tau_0\right)}{\partial\tau_0}\right)\\
    &:={\bm{S}'\left(\tau_0\right)}^{\mathcal{H}}\text{diag}\left(\bm{S}_{\bm{w}}\right)^{-1}{\bm{S}'\left(\tau_0\right)}.
  \end{split}
\end{equation}

For the sensing subsystem, the $n$-th elements of $\bm{s}\left(\tau_0\right)$ and $\bm{w}$ are defined respectively as
\begin{subequations}
  \begin{align}
    s\left(n;\tau_0\right)=&\mathfrak{R}\mathcal{K}x_s\left(\frac{n}{R_s}-\tau_0\right),\\
    w\left(n\right)=&\mathfrak{R}w_{p}\left(n\right)+w_s\left(n\right).
  \end{align}
\end{subequations}

Therefore, the $k$-th element of $\bm{S}'\left(\tau_0\right)$ is derived as
\begin{equation}
  \begin{split}
    &S'\left(k;\tau_0\right)=\frac{1}{\sqrt{N}}\sum_{n=0}^{N-1}{\frac{\partial s\left(n;\tau_0\right)}{\partial\tau_0}\exp\left(-\frac{j2\pi nk}{N}\right)}\\
    &=\frac{1}{\sqrt{N}}\sum_{n=0}^{N-1}{\left(-\mathfrak{R}\mathcal{K}\frac{dx_s\left(t\right)}{dt}|_{t=\frac{n}{R_s}-\tau_0}\right)\exp\left(-\frac{j2\pi nk}{N}\right)}\\
    &=-\mathfrak{R}\mathcal{K}\left(j2\pi k\Delta f\right)\tilde{H}_s\left(k\right)X\left(k\right)\exp\left(-j2\pi k\Delta f\tau_0\right).
  \end{split}
\end{equation}

Consequently, the Fisher information $I$ is derived as
\begin{equation}\label{Fisher_I_exp}
  \begin{split}
    I&=\sum_{k=0}^{N-1}{\frac{{\lvert S'\left(k;\tau_0\right) \rvert}^2}{S_{\bm{w}}\left(k\right)}}\\
    &=4\pi^2{\Delta f}^2\sum_{k=0}^{N-1}{\dfrac{k^2\mathfrak{R}^2\mathcal{K}^2{\lvert \tilde{H}_s\left(k\right) \rvert}^2\mathbb{E}\left({\lvert X\left(k\right) \rvert}^2\right)}{N\left(\mathfrak{R}^2{\lvert \tilde{H}_s\left(k\right) \rvert}^2P_{w_{p}}\left(k\right)+\dfrac{N_s\Delta f}{2\mathbb{E}\left(\bar{h}_s^2\right)}\right)}}\\
    &=\frac{8\pi^2{\Delta f}^2}{N}\sum_{k=0}^{\frac{N}{2}-1}{\dfrac{k^2\mathcal{K}^2{\lvert \tilde{H}_s\left(k\right) \rvert}^2\left(P-b^2\right)\tilde{P}\left(k\right)}{\left({\lvert \tilde{H}_s\left(k\right) \rvert}^2P_{w_{p}}\left(k\right)+\dfrac{N_s\Delta f}{2\mathfrak{R}^2\mathbb{E}\left(\bar{h}_s^2\right)}\right)}}\\
    &=\frac{8\pi^2{\Delta f}^2}{N}\sum_{k=0}^{\frac{N}{2}-1}{k^2\gamma_s\left(k\right)\tilde{P}\left(k\right)}.
  \end{split}
\end{equation}


\section{Proof of \textbf{Theorem 2}}\label{append_clipping_corr}
Regarding the clipping noise $w_p\left(n\right)$ as the output of a memoryless non-linear system $g_p\left(x\left(n\right)\right)$, the following \textit{Lemma 3} can be adopted.

\textit{Lemma 3 (Price theorem): }Consider a bivariate normal distribution in variables $U,V$ with covariance $r=\mathbb{E}\left(UV\right)-\mathbb{E}\left(U\right)\mathbb{E}\left(V\right)$ and a regularized function $g\left(u,v\right)$. Then, the expectation of a random variable $g\left(U,V\right)$ has the property of
\begin{equation}
  \frac{\partial^n\mathbb{E}\left(g\left(U,V\right)\right)}{\partial r^n}=\mathbb{E}\left(\frac{\partial^{2n}g\left(U,V\right)}{\partial U^n\partial V^n}\right).
\end{equation}

\textit{Proof: }\textit{Lemma 3} can be derived by the characteristic function of a bivariate Gaussian distribution. Please see~\cite{Price_Theorem} for more details.

If we set the regularized function in \textit{Lemma 3} as $g\left(u,v\right)=g_{p}\left(u\right)g_{p}\left(v\right)$, then the auto-correlation function of the clipping noise has the property of
\begin{equation}
  \begin{split}
    \frac{\partial^2R_{w_{p}}\left(n-m\right)}{\partial r^2}&=\frac{\partial^2\mathbb{E}\left(g\left(x\left(m\right),x\left(n\right)\right)\right)}{\partial r^2}\\
    &=\mathbb{E}\left(\frac{\partial^4g\left(x\left(m\right),x\left(n\right)\right)}{\partial{x\left(m\right)}^2\partial{x\left(n\right)}^2}\right)\\
    &=\mathbb{E}\left(\frac{\partial^2g_{p}\left(x\left(m\right)\right)}{\partial{x\left(m\right)}^2}\frac{\partial^2g_{p}\left(x\left(n\right)\right)}{\partial{x\left(n\right)}^2}\right)\\
    &=\mathbb{E}\left(\delta\left(x\left(m\right)+b\right)\delta\left(x\left(n\right)+b\right)\right)\\
    &=\frac{1}{2\pi\sqrt{\sigma_x^4-r^2}}\exp\left(-\frac{b^2}{\sigma_x^2+r}\right),
  \end{split}
\end{equation}

{\noindent}where $r$ denotes the auto-correlation function of $x\left(n\right)$, and the notation $\delta\left(x\right)$ is the Dirac delta function.

Calculating the indefinite integral yields the auto-correlation function of $w_{p}\left(n\right)$ as
\begin{equation}
  R_{w_{p}}\left(n\right)=\mathcal{I}\left(r\right)+C_1r+C_2,
\end{equation}

{\noindent}where the integral $\mathcal{I}\left(r\right)$ is defined in (\ref{Price_integral}), while $C_1$ and $C_2$ are constants independent from $r$. Then, by considering two special values of $r$, the constants $C_1$ and $C_2$ can be attained.

When $r=R_x\left(n-m\right)=0$ for $n\neq m$, i.e., $x\left(m\right)$ and $x\left(n\right)$ are independent statistically, the auto-correlation function of $w_{p}\left(n\right)$ is given by
\begin{equation}\label{C1_eq}
  R_{w_{p}}\left(n\right)=\left(\mathbb{E}\left(w_{p}\left(n\right)\right)\right)^2=\mathcal{I}\left(0\right)+C_2.
\end{equation}

When $r=R_x\left(n-m\right)=\sigma_x^2$ for $\forall n,m$, i.e., $x\left(m\right)$ always equals $x\left(n\right)$, the auto-correlation function of $w_{p}\left(n\right)$ is given by
\begin{equation}\label{C2_eq}
  R_{w_{p}}\left(n\right)=\mathbb{E}\left(\left(w_{p}\left(n\right)\right)^2\right)=\mathcal{I}\left(\sigma_x^2\right)+C_1\sigma_x^2+C_2.
\end{equation}

Consequently, by simultaneously solving (\ref{C1_eq}) and (\ref{C2_eq}), the expressions of (\ref{Ya}) and (\ref{Yb}) can be obtained.

\section{Proof of \textit{Lemma 1}}\label{append_convergence_exist}
As indicated by (\ref{xi_1_definition}) and (\ref{xi_2_definition}), $\xi_1\left(\mu,\eta\right)$ and $\xi_2\left(\mu,\eta\right)$ are both continuous, non-increasing with respect to $\mu$, and non-decreasing with respect to $\eta$ in the first quadrant. In addition, the existence of $\varGamma_1$ and $\varGamma_2$ is guaranteed by
\begin{subequations}
  \begin{align}
    &\lim_{\mu\to\infty}& &{\xi_1\left(\mu,\eta\right)}=0,\\
    &\lim_{\mu\to 0}& &{\xi_1\left(\mu,\eta\right)}=\left(\frac{N}{2}-1\right)\tilde{P}_m\geq\frac{1}{2},\\
    &\lim_{\mu\to\infty}& &{\xi_2\left(\mu,\eta\right)}=0,\\
    &\lim_{\mu\to 0}& &{\xi_2\left(\mu,\eta\right)}=\sum_{k=1}^{\frac{N}{2}-1}{k^2\gamma_s\left(k\right)\tilde{P}_m}\geq\tilde{\varsigma}_0^2.
  \end{align}
\end{subequations}

In consequence, as contour lines of $\xi_1\left(\mu,\eta\right)$ and $\xi_2\left(\mu,\eta\right)$, curves $\varGamma_1$ and $\varGamma_2$ have direction vectors pointing to the first quadrant. Without loss of generality, they are rewritten as $\varGamma_1:\ \mu=\zeta_1\left(\eta\right)$ and $\varGamma_2:\ \mu=\zeta_2\left(\eta\right)$, where $\zeta_1\left(\eta\right)$ and $\zeta_2\left(\eta\right)$ are continuous and non-decreasing with respect to $\eta$.

Moreover, if we set $\eta=0$, $\zeta_1\left(0\right)$ corresponds to $\tilde{P}_c\left(k\right)$ as described in (\ref{opt_Pk_c_only}). However, considering the assumption of \textit{Case C}, the normalized power allocation for optimal communication performance cannot satisfy the sensing CRB constraint, i.e.,
\begin{equation}
  \xi_2\left(\zeta_1\left(0\right),0\right)\leq\tilde{\varsigma}_0^2.
\end{equation}

{\noindent}Therefore, $\zeta_1\left(0\right)\geq\zeta_2\left(0\right)$ can be derived as $\xi_2\left(\mu,0\right)$ is non-increasing to $\mu$.

On the other hand, for a sufficiently large $\eta^+$, $\zeta_1\left(\eta^+\right)$ corresponds to $\tilde{P}_s\left(k\right)$, and the assumption of \textit{Case C} reveals its feasibility, i.e.,
\begin{equation}
  \xi_2\left(\zeta_1\left(\eta^+\right),0\right)\geq\tilde{\varsigma}_0^2,
\end{equation}

{\noindent}which further indicates that $\zeta_1\left(\eta^+\right)\leq\zeta_2\left(\eta^+\right)$. Consequently, the Intermediate Value theorem gives that $\exists\ \eta^*>0,\zeta_1\left(\eta^*\right)=\zeta_2\left(\eta^*\right)$, which completes the proof of \textit{Lemma 1}~\cite{hughes_calculus}.

\section{Proof of \textit{Lemma 2}}\label{append_convergence_region}
\textit{Lemma 2} can be proven by contradiction. Supposing that $\mu<\gamma_s\left(l_2\right)l_2^2\eta$, then the power allocation has the property of
\begin{subequations}
  \begin{align}
    &\xi_0\left(\mu,\eta,k\right)=\dfrac{1}{\tilde{P}_m+1/\gamma_c\left(k\right)},\ \forall k,\\
    &\xi_1\left(\mu,\eta\right)=\left(\frac{N}{2}-1\right)\tilde{P}_m>\frac{1}{2}.
  \end{align}
\end{subequations}

{\noindent}However, the total-power constraint is violated, and therefore the assumption is invalid.

Similarly, supposing that
\begin{equation}
  \mu>\gamma_s\left(l_1\left(\eta\right)\right)l_1\left(\eta\right)^2\eta+\gamma_c\left(l_1\left(\eta\right)\right),
\end{equation}

{\noindent}then the power allocation has the property of
\begin{subequations}
  \begin{align}
    &\xi_0\left(\mu,\eta,k\right)=\mu-\gamma_s\left(k\right)k^2\eta>\gamma_c\left(k\right),\forall k\\
    &\xi_1\left(\mu,\eta\right)=\sum_{k=1}^{\frac{N}{2}-1}{\left\{\frac{1}{\mu-\gamma_s\left(k\right)k^2\eta}-\frac{1}{\gamma_c\left(k\right)}\right\}^+}=0.
  \end{align}
\end{subequations}

Obviously, $\left(\mu,\eta\right)$ cannot be the solution to (P1-2) neither, and therefore the assumption is also invalid, which completes the proof of \textit{Lemma 2}.
}

\bibliographystyle{IEEEtran}
\bibliography{Ref.bib}

\vfill

\end{document}